\documentclass[aps,preprint,a4paper, showkeys]{revtex4-1}
\usepackage{graphicx}
\usepackage{amsmath}
\usepackage{amssymb}
\usepackage{times}
\usepackage[utf8]{inputenc}
\usepackage{color}
\usepackage{ulem}
\usepackage[pdfborder=0]{hyperref}
\hypersetup{
	colorlinks=true,
    linkcolor=blue,          % color of internal links
    citecolor=blue			 % color of citations
}
\def\nn{\nonumber}
\def\beq{\begin{equation}}
\def\eeq{\end{equation}}
\def\beqna{\begin{eqnarray}}
\def\eeqna{\end{eqnarray}}
\def\bea{\begin{array}}
\def\ea{\end{array}}

\def\MU{{\mathcal U}}
\def\MV{{\mathcal V}}
\def\mw{{\mathcal W}}

\begin{document}
\title{Heating and thermal squeezing in parametrically-driven oscillators
with added noise}
\author{Adriano A. Batista}
\affiliation{
Departamento de Física\\
Universidade Federal de Campina Grande\\
Campina Grande-PB\\
CEP: 58109-970\\
Brazil}
\date{\today}
\begin{abstract}
In this paper we report a theoretical model based on Green’s
functions, Floquet theory and averaging techniques up to second order that
describes the dynamics of parametrically-driven oscillators with added thermal
noise. Quantitative estimates for heating and quadrature thermal noise squeezing
near and below the transition line of the first parametric instability zone of
the oscillator are given. Furthermore, we give an intuitive explanation as to why
heating and thermal squeezing occur.
For small amplitudes of the parametric pump the Floquet multipliers are complex
conjugate of each other with a constant magnitude. As the pump amplitude is
increased past a threshold value in the stable zone near the first parametric instability, the
two Floquet multipliers become real and have different magnitudes.
This creates two different effective dissipation rates (one smaller and
the other larger than the real dissipation rate) along the stable manifolds of
the first-return Poincaré map.
We also show that the statistical average of the input power due to thermal
noise is constant and independent of the pump amplitude and frequency. 
The combination of these effects cause most of heating and thermal
squeezing. Very good agreement between analytical and numerical estimates of the thermal
fluctuations is achieved.

\end{abstract}
\maketitle
\section{Introduction}
%Background on parametric systems
Parametrically-driven systems and parametric resonance occur in many different
physical systems, ranging from the mechanical domain to the electronic,
microwave, electromechanic, optomechanic, and quantum domains. 
In the mechanical domain we have Faraday waves \cite{faraday1831},
inverted pendulum stabilization, stability of boats, balloons, and parachutes \cite{ruby1996}. 
A comprehensive review of applications in electronics and microwave cavities
spanning from the early twentieth century up to 1960 can be found in
Ref.~\cite{mumford1960}. 
A few relevant recent applications, in
micro and nano systems, include quadrupole ion guides and ion traps \cite{paul90},
linear ion crystals in linear Paul traps designed as
prototype systems for the implementation of quantum computing
\cite{raizen1992ionic, drewsen1998large, kielpinski2002architecture}, 
magnetic resonance force microscopy \cite{dougherty1996detection}, 
tapping-mode force microscopy \cite{moreno2006parametric}, axially-loaded
microelectromechanical systems (MEMS) \cite{requa06electromechanical},
torsional MEMS \cite{turner98nature}.
In the quantum domain we could mention  wideband superconducting parametric
amplifiers \cite{eom2012nature} and squeezing
in optomechanical cavities below the zero-point motion \cite{szorkovszky2011}.

%review of related bibliography in classical thermal squeezing
Parametric pumping has had many applications in the field of MEMS, which have
been used primarily as accelerometers, for measuring small forces and as ultrasensitive mass detectors
since the mid 80's \cite{binnig86}. 
An enhancement to the detection techniques in MEMS was developed by Rugar and
Grütter \cite{rugar91} in the early 90's that uses mechanical parametric amplification (before transduction) 
to improve the sensitivity of measurements.  
This amplification method works by driving the parametrically-driven resonator
on the verge of parametric unstable zones. 
%Description of related experiments
They were
looking for means of reducing noise and increasing precision in a detector
for gravitational waves, when they experimentally found classical
thermomechanical quadrature squeezing, a phenomenon which is reminiscent of quantum squeezed states.
The classical version is characterized
by oscillating levels of the response of the parametric oscillator to noise at
the frequency of the parametric pump, in such a way that the product between
maximum and minimum output noise levels is constant. They observed that when
the pump was turned on, the noise increased in one quadrature, while on the
other it decreased. No theoretical model was proposed by them to explain the effect though. 
Subsequently, DiFilippo et al. \cite{difilippo92prl}
and Natarajan et al. \cite{natarajan95prl} proposed
theoretical explanations for this noise squeezing phenomenon, but their models
did not treat noise directly in the equations of motion.

%Description of related experiments

% Description of other theoretical models 

% What we do here
Here, we study  a parametrically-driven
oscillator in the presence of noise with the objective of understanding what
causes thermal squeezing and heating in the stable zone near the transition
line of the first parametric instability. 
The one-degree of freedom model studied here may be applied for instance to the
fundamental mode of a doubly-clamped beam resonator that is axially loaded, in which case 
the one degree of freedom represents the amount of deflexion of the middle of
the beam from the equilibrium position. The present model can also be applied 
to the linear response of ac driven nonlinear oscillators to noise (such as
transversally-loaded beam resonators), see for example Ref.~\cite{almog2007noise}. 

% Statement of our objectives
One of the objectives of the present investigation is to extend and improve on
recently obtained analytical quantitative estimates of the amount of quadrature noise squeezing 
and heating in a parametrically-driven oscillator \cite{batista2011cooling}.
%
% Description of our theoretical model+achievements
% How we do it
Here we use the Green's function approach, previously developed to solve the
Langevin equation, aligned with averaging techniques up to second order, to
obtain more precise analytical estimates of the thermal fluctuations in the
parametrically-driven oscillator with added noise.
We further show, using an approximate Floquet theory based on first and
second-order averaging approximations, that thermal squeezing and heating are
related to the onset of real-valued Floquet multipliers (FMs) with different
magnitudes. It is shown that one FM grows while one gets closer in parameter
space to the first transition line to instability while the other FM decreases. 
As a consequence, one gets two different effective dissipation rates,
while at the same time the input power due to noise remains constant as the pump
amplitude is increased. 
We show below that these effects account for most thermal squeezing and heating
observed. 
Furthermore, first-order analytical estimates of heating and the amount of
squeezing are also provided.
% Motivation, why we do it

The contents of this paper are organized as follows. In Sec.~(\ref{theory})
we present our theoretical model, in Sec.~(\ref{numerics}) we present and
discuss our numerical results, and in Sec.~(\ref{conclusions}) we draw our
conclusions.
\section{Theory}
\label{theory}
\noindent
The equation for the parametrically-driven  oscillator (in dimensionless format) is given by
the damped Matthieu's equation
\begin{equation}
    \ddot x+\omega^2_0x=-\gamma\dot x+F_p\cos(2\omega t)\;x,
    \label{parOsc}
\end{equation}
in which $\gamma$ and $F_p\sim O(\varepsilon)$, where $\varepsilon<<1$.
Since we want to apply the averaging method (AM) \cite{verh96, Guck83} to
situations in which we have detuning, it is convenient to rewrite
Eq.~(\ref{parOsc}) in a more appropriate form with the
notation $\Omega=\omega^2_0-\omega^2$, where we also have $\Omega\sim O(\varepsilon)$.
With this substitution we obtain
$\ddot x+\omega^2x=-\Omega x-\gamma\dot x+F_p\cos(2\omega t)\;x$.
We then rewrite this equation in the form
$\dot x = y$, $\dot y = -\omega^2x+f(x,y,t)$,
where $ f(x,y,t)= -\Omega x+F_p\cos(2\omega t)\;x-\gamma y$.
We now set the above equation in slowly-varying form with the transformation to a slowly-varying frame
\beq
\left(\bea{c}
x \\ y
\ea
\right)
=
\left(
\begin{array}{cc}
    \cos\omega t & -\sin \omega t\\
    -\omega\sin\omega t & -\omega\cos\omega t
\end{array}
\right)
\left(\bea{c}
\MU \\ \MV
\ea
\right)
\label{slow_trans}
\eeq
   and obtain 
\beqna
\left(\bea{c}
\dot \MU \\ \dot \MV
\ea
\right)
&=&
\left(
\begin{array}{cc}
    \cos \omega t & -\frac{1}{\omega}\sin\omega t\\
    -\sin\omega t & -\frac{1}{\omega}\cos\omega t\nn
\end{array}
\right)
\left(\bea{c}
0\\  f(x,y,t)
\ea
\right)\nn\\
&=&
-\frac{1}{\omega}\left(\bea{c} \sin\omega t f(x,y,t)\\\cos\omega t
f(x,y,t)\ea\right)\equiv F(\MU, \MV, t)=DF(\MU, \MV, t)\left(\bea{c}
\MU \\ \MV
\ea
\right).
\label{slow_frame}
\eeqna

The components of the jacobian matrix $DF(\MU, \MV, t)$ of the above flow is
given by
\beqna
DF_{11} &=&
\frac{-1}{2\omega}\left(
\gamma\omega  (1-\cos(2\omega
    t))-\Omega\sin(2\omega t) +\frac{F_p}{2}\sin(4\omega t)\right)\nn\\
DF_{12} &=&\frac{-1}{2\omega}\left(
    \Omega(1-\cos(2\omega t)) +\frac{F_p}{2}(1-2\cos(2\omega
    t)+\cos(4\omega t))+\gamma\omega\sin(2\omega t)\right)\nn\\ 
DF_{21} &=& \frac{-1}{2\omega}\left(
    -\Omega(1+\cos(2\omega t))
    +\frac{F_p}{2}[1+2\cos(2\omega t)+\cos(4\omega t)]+\gamma\omega \sin(2\omega
    t)\right)\nn\\
DF_{22} &=&\frac{-1}{2\omega}\left(\Omega\sin(2\omega
    t)+\gamma\omega\left(1+\cos(2\omega t)\right)-\frac{F_p}{2}\sin(4\omega t)
    \right)
\eeqna
After application of the AM to first order (in which, basically,
we filter out oscillating terms at $2\omega$ and $4\omega$ in the above
equation), we obtain
\begin{eqnarray}
    \dot { u} &=& \frac{-1}{2\omega}\left[\gamma\omega  u
    +\left(\Omega +\frac{F_p}{2}\right) v\right],\nn\\
    \dot { v} &=& \frac{-1}{2\omega}\left[\left(-\Omega +\frac{F_p}{2}\right) u+\gamma\omega v\right],
    \label{1stOrdAvg}
\end{eqnarray}
where the functions $\MU(t)$ and $\MV(t)$ are related to their slowly-varying
averages $u(t)$ and $v(t)$, respectively, by the transformation
\beq
\left(\bea{c}
\MU \\ \MV
\ea
\right)=\left(\bea{c}
 u +\mw_1\\  v+\mw_2
\ea
\right).
\eeq
According to the averaging theorem \cite{hol81, bat00}, the vector
$\mw$ obeys $\partial_t\mw(u,v,t)=g (u,v,t)$, where the vector $g(u, v, t)$
corresponds to the explicitly-varying components of the right side of
Eq.~(\ref{slow_frame}). 
Namely, we have
{
\beqna
&&\frac{\partial\mw(u,v, t)}{\partial
t}=
\left( 
\bea{c}
g_1 (u, v, t) \\
g_2 (u,v, t)
\ea 
\right)=\nn\\
&=&\frac{1}{2\omega}\left( 
\bea{c}
\left[(\Omega+F_p){v}+\omega\gamma{u}\right]\cos(2\omega
t)
+\left(\Omega{u}-\gamma\omega{v}\right)\sin(2\omega
t)-\frac{F_p}{2}\left[{v}\cos(4\omega
t)+{u}\sin(4\omega t)\right]\\ 
\left[(\Omega-F_p){u}-\omega\gamma{v}\right]\cos(2\omega
t)
-\left(\Omega{v}+\gamma\omega{u}\right)\sin(2\omega
t)-\frac{F_p}{2}\left[{u}\cos(4\omega t)-{v}\sin(4\omega
t)\right]
\ea 
\right).\nn
\eeqna
}
Upon integration we find 

\begin{flushleft}
$\mw(u,v, t)=\frac{1}{4\omega^2}\times$
\[
\left[
\bea{ll}
\omega\gamma\sin(2\omega t)-\Omega\cos(2\omega t)+\frac{F_p}{4}\cos(4\omega t)
&
(\Omega+F_p)\sin(2\omega t)+\gamma\omega\cos(2\omega
t)-\frac{F_p}{4}\sin(4\omega t)
\\
(\Omega-F_p)\sin(2\omega t)+\gamma\omega\cos(2\omega
t)-\frac{F_p}{4}\sin(4\omega t) & -\omega\gamma\sin(2\omega t)+\Omega\cos(2\omega t)-\frac{F_p}{4}\cos(4\omega t)
\ea 
\right]\]
$\times\left[\bea{c} u \\ v\ea\right],$
\end{flushleft}
in which the integration constants are set to zero.
The averaging theorem \cite{Guck83} states that
these two sets of functions, namely $(u(t), v(t))$  and $(\MU(t),\, \MV(t))$,
will be close to each other to order $O(\epsilon)$ during a time scale of $O(1/\epsilon)$ if they have initial conditions within an initial distance of $O(\epsilon)$.
So by studying the simpler averaged system, one may obtain very accurate
information about the corresponding more complex non-autonomous original system.
Using the transformations $ u(t)= e^{-\gamma t/2}\tilde u(t)$ and $v(t)=
e^{-\gamma t/2}\tilde v(t)$ in Eqs.~\eqref{1stOrdAvg}, we obtain
\begin{eqnarray}
    \dot{\tilde u} &=& \frac{-1}{2\omega}\left(\Omega+\frac{F_p}{2}\right)\tilde v,\nn\\ 
    \dot{\tilde v} &=&
    \frac{-1}{2\omega}\left(-\Omega+\frac{F_p}{2}\right)\tilde u.
    \label{eq_av1}
\end{eqnarray}
Upon integration of
Eqs.~(\ref{eq_av1}), one finds the solution
\beqna
u(t)&=&e^{-\gamma t/2}\left[u_0 \cosh (\kappa
t)+\frac{\beta -\delta}{\kappa}v_0\sinh(\kappa t)\right]\nn,\\
v(t)&=&e^{-\gamma t/2}\left[v_0 \cosh (\kappa
t)+\frac{\beta+\delta}{\kappa}u_0 \sinh(\kappa t)\right],
\label{1st-avg_sol}
\eeqna
where $\kappa=\sqrt{\beta^2-\delta^2}$, $\beta=-F_p/4\omega$, and 
$\delta=\Omega/2\omega$. Hence, we find that the first parametric resonance, i.e. the boundary between
the stable and unstable responses, is given by 
\beq
(\gamma\omega)^2=(F_p/2)^2-\Omega^2.
\label{eq_zones}
\eeq
This result is valid for $\omega\approx \omega_0$ even in the presence of
added noise. In Fig.~\ref{zones} we find very good agreement between the
boundary obtained from numerical integration of Eq.~(\ref{parOsc}) and the boundary given by the
averaging technique.

From Eqs. (\ref{slow_trans}) and (\ref{1st-avg_sol}) we obtain the approximate
fundamental matrix $\Phi(t)$ (also known as the time evolution operator)

\beqna
\left(\bea{c}
x(t) \\ y(t)
\ea
\right)
&=&\Phi(t)\left(\bea{c}
x_0 \\ y_0
\ea
\right)\nn\\
&\approx&
e^{-\gamma t/2}\times\left(
\begin{array}{cc}
    \cos\omega t & -\sin \omega t\\
    -\omega\sin\omega t & -\omega\cos\omega t
\end{array}
\right)
\left(\bea{cc}
\cosh(\kappa\,t)  & -\frac{\beta-\delta}{\omega\kappa}\sinh(\kappa\,t)\\
\frac{\beta+\delta}{\kappa}\sinh(\kappa\,t)&-\frac{1}{\omega}\cosh(\kappa\,t) \\
\ea
\right)\left(\bea{c}
x_0 \\ y_0
\ea
\right)\nn,
\eeqna
where $\Phi(0)=I$, in which $I$ is the identity matrix.
From Floquet theory \cite{verh96} we know that $\Phi(t)=P(t)e^{Bt}$, where 
$P(t)$ is a periodic matrix with period $T=\pi/\omega$. We also know that $P(0)=I$.
The eigenvalues of $e^{BT}$ are known as the Floquet multipliers.
We rewrite the fundamental matrix in the following form
\beq
\Phi(t)
\approx
\left(
\begin{array}{cc}
    \cos\omega t & \frac{1}{\omega}\sin \omega t\\
    -\omega\sin\omega t & \cos\omega t
\end{array}
\right)
\left(\bea{cc}
e^{-\gamma t/2}\cosh(\kappa\,t)  &
-\frac{\beta-\delta}{\omega\kappa}e^{-\gamma t/2}\sinh(\kappa\,t)\\
-\frac{\omega(\beta+\delta)}{\kappa}e^{-\gamma
t/2}\sinh(\kappa\,t)&e^{-\gamma t/2}\cosh(\kappa\,t) \\
\ea
\right).
\eeq
Hence, we notice that from the approximate solution of the fundamental matrix,
via first-order averaging, we can find the approximate Floquet multipliers.
They are given by
\beq
\lambda_\pm=-e^{-(\frac{\gamma}{2}\pm\kappa)T}.
\eeq

Further improvements can be made by going to second order averaging.
According to Ref.~\cite{bat00},
the second-order corrections are given by the time average of

\beqna
&&\overline{DF(\MU, \MV, t)\mw(\MU, \MV)}=\nn\\
&=&\frac{-1}{8\omega^3}\left(
\bea{cc}
-\gamma\omega F_p& -\gamma^2\omega^2-\Omega(\Omega+ F_p)-F_p^2/8\\
\gamma^2\omega^2+\Omega(\Omega-F_p) +F_p^2/8& \gamma \omega F_p
\ea
\right)
\left(
\bea{c}
u \\ v
\ea
\right).
\nn
\eeqna
Hence, the second-order approximation to Eq.~\eqref{slow_frame} becomes
\begin{eqnarray}
    \dot { u} &=& \frac{-1}{2\omega}\left[\gamma(\omega - \frac{F_p}{4\omega}) u
    +\left(\Omega+\frac{F_p}{2}-\frac{\gamma^2}{4}-\frac{\Omega(\Omega+F_p)}{4\omega^2}-\frac{F_p^2}{32\omega^2}\right)
    v\right],\nn\\ 
    \dot { v} &=& \frac{-1}{2\omega}\left[\left(-\Omega
    +\frac{F_p}{2}+\frac{\gamma^2}{4}+\frac{\Omega(\Omega-F_p)}{4\omega^2}+\frac{F_p^2}{32\omega^2}\right)
    u+\gamma(\omega +\frac{F_p}{4\omega})v\right].
    \label{eq_av2}
\end{eqnarray}
The solution is given by
\beqna
u(t) &=& e^{-\gamma t/2}\left[\left(
\cosh(\xi t)+a\frac{\sinh(\xi t)}{\xi}\right)u_0+(b-c)\frac{\sinh(\xi t)}{\xi}
v_0\right],\nn\\ 
v(t) &=& e^{-\gamma t/2}\left[(b+c)\frac{\sinh(\xi t)}{\xi} u_0+\left(
\cosh(\xi t)-a\frac{\sinh(\xi t)}{\xi}\right)v_0\right],
\eeqna
with $\xi=\sqrt{a^2+b^2-c^2}$, $a=-\frac{\gamma\beta}{2\omega}$,
$b=\beta-\delta\beta/\omega$, and
$c=\delta-\gamma^2/(8\omega)-\delta^2/(2\omega)-\beta^2/(4\omega).$
Thus, We can write the fundamental matrix in second-order approximation as 
\beq
\Phi(t)
\approx
e^{-\gamma t/2}\left(
\begin{array}{cc}
    \cos\omega t & \frac{1}{\omega}\sin \omega t\\
    -\omega\sin\omega t & \cos\omega t
\end{array}
\right)
\left(\bea{cc}
\cosh(\xi\,t) +a\frac{\sinh(\xi\,t)}{\xi} &
-\frac{b-c}{\omega\xi}\sinh(\xi\,t)\\
-\frac{\omega(b+c)}{\xi}\sinh(\xi\,t)&\cosh(\xi\,t)-a\frac{\sinh(\xi\,t)}{\xi}
\ea
\right).
\nn
\eeq
After some simple algebraic operations, we find the Floquet multipliers
(eigenvalues of $\Phi(T)$) to be given by
\beq
\lambda_\pm=-e^{-(\gamma/2\pm\xi )T}.
\label{2ndOrderFM}
\eeq
Hence, the transition line to instability in second-order averaging is
given by 
\beq
\gamma=2\xi.
\label{2ndOrderTransLine}
\eeq
 
\subsection{Green's function method}
\noindent
The equation for the Green's function of the parametrically-driven oscillator is
given by
\begin{equation}
    \left[\frac{\partial^2}{\partial t^2} +\omega_0^2+\gamma \frac{\partial}{\partial t}-F_p\cos(2\omega t)\right]\;G(t,t')=\delta(t-t').
    \label{green_eq}
\end{equation}
Since we are interested in the stable zones of the parametric oscillator,
for $t<t'$ $G(t,t')=0$ and by integrating the above equation near $t=t'$, 
we obtain the initial conditions when $t=t'+0^+$, $G(t,t')=0$ and $\frac{\partial}{\partial t}G(t,t')=1.0$.
\subsubsection{1st-order averaging}
Although Eq. \eqref{green_eq} may be solved exactly by using Floquet
theory \cite{wies85}, one obtains very complex
 solutions.
Instead, we find fairly simple analytical approximations to the Green's
functions and, subsequently, to the statistical averages of fluctuations using
the averaging method. 
From Eq.~\eqref{slow_trans} we obtain the approximate Green's function is
$G(t,t')= \cos(\omega t)\MU(t)-\sin(\omega t)\MV(t)= \cos(\omega t)[u(t)+\mw_1(u(t), v(t), t)]-\sin(\omega t)[(v(t)+\mw_2(u(t), v(t),
t)]$, the functions $u(t)$ and $v(t)$ are given by the solution of the system of
coupled differential equations \eqref{1st-avg_sol}, where the time $t$ is
replaced by $t-t'$ and the initial conditions set at $t=t'$ are given by
$u(t')=-\sin(\omega t')/\omega$ and $v(t')=-\cos(\omega t')/\omega$. 
For simplicity we set $\mw_1(t')=\mw_2(t')=0$. 
We find the approximate Green's function to be
\beqna
G(t,t')&\approx& -\frac{e^{-\gamma (t-t')/2}}{\omega}\left[\cos(\omega
t)\left(\cosh(\kappa\,(t-t'))\sin(\omega t')+\frac{\beta-\delta}{\kappa}\sinh(\kappa\,(t-t'))\cos(\omega t')\right)\right.\nn\\ &&\left.-\sin(\omega
t)\left(\frac{\beta+\delta}{\kappa}\sinh(\kappa\,(t-t'))\sin(\omega
t')+\cosh(\kappa\,(t-t')) \cos(\omega t')\right)\right]\nn\\
&=&
\frac{e^{-\gamma (t-t')/2}}{\omega} \left\{
\cosh(\kappa\,(t-t'))\sin(\omega (t-t'))
+\frac{\delta}{\kappa}\sinh(\kappa\,(t-t'))\cos(\omega (t-t'))\right.\nn\\
&&\left.-\frac{\beta}{\kappa}\sinh(\kappa\,(t-t'))\cos(\omega
(t+t')) \right\},
\label{green_avg}
\eeqna 
for $t>t'$ and $G(t,t')=0$ for $t<t'$. In the stable zone of the
parametrically-driven oscillator, when $|\beta|>|\delta|$ , we can rewrite the
Green's function replacing the initial conditions and using simplifying
trigonometrical identities. The change of variables $t'=t-\tau$ leads to 
\beqna
G(t,t-\tau)&\approx& \frac{e^{-\gamma \tau/2}}{\omega} \left\{
\cosh(\kappa\,\tau)\sin(\omega \tau)
+\frac{\delta}{\kappa}\sinh(\kappa\,\tau)\cos(\omega \tau)\right.\nn\\
&&\left.-\frac{\beta}{\kappa}\sinh(\kappa\,\tau)\cos(\omega (2t-\tau)) \right\}.
\label{green_avg1}
\eeqna
\subsubsection{2nd-order averaging}
\label{2nd-orderGF}
The Green's function obtained by second-order averaging is given by
\beqna
&&G(t,t')\approx -\frac{e^{-\gamma
(t-t')/2}}{\omega}\times\nn\\
&&\left\{\cos(\omega t) \left[\left(
\cosh(\xi (t-t'))+a\frac{\sinh(\xi
(t-t'))}{\xi}\right)\sin(\omega t')+(b-c)\frac{\sinh(\xi (t-t'))}{\xi}
\cos(\omega t')\right]\right.\nn\\ 
&&\left.-\sin(\omega t)
\left[(b+c)\frac{\sinh(\xi (t-t'))}{\xi} \sin(\omega t')+\left( \cosh(\xi (t-t'))-a\frac{\sinh(\xi
(t-t'))}{\xi}\right)\cos(\omega t')\right]\right\}\nn,
\eeqna 
for $t>t'$ and $G(t,t')=0$ for $t<t'$.
This can be rewritten in a shorter format as
\beqna
&&G(t,t-\tau)\approx \frac{e^{-\gamma\tau/2}}{\omega}\left\{
\cosh(\xi \tau)\sin(\omega\tau) +c\frac{\sinh(\xi \tau)}{\xi}
\cos(\omega\tau)\right.\nn\\ 
&&\left.-\frac{\sinh(\xi\tau)}{\xi}\left[a\sin(\omega
(2t-\tau))+b\cos(\omega (2 t-\tau)\right]\right\},
\label{green_avg2}
\eeqna 
for $\tau>0$ and $G(t,t-\tau)=0$ for $\tau<0$.
We notice that this second-order approximate Green's function can be put in the
same format as the one of the first-order approximation given by
Eq.~\eqref{green_avg1}.
\beqna
&&G(t,t-\tau)\approx \frac{e^{-\gamma\tau/2}}{\omega}\left\{
\cosh(\xi \tau)\sin(\omega\tau) +c\frac{\sinh(\xi \tau)}{\xi}
\cos(\omega\tau)\right.\nn\\ 
&&\left.-\beta'\;\frac{\sinh(\xi\tau)}{\xi}\cos(\omega[2(t-t_0)-\tau])\right\},
\label{green_avg2n}
\eeqna 
where $\beta'= \beta\sqrt{(1-\delta/\omega)^2+\gamma^2/(4\omega^2)}$,
$a=\beta'\sin(2\omega t_0)$, and $b=\beta'\cos(2\omega t_0)$. 
In Eq.~\eqref{green_avg1} we have to replace $\kappa$ by $\xi$, $\delta$ by
$c$, $\beta$ by $\beta'$ and $t$ by $t-t_0$ to obtain the second-order
Green's function given by Eq.~\eqref{green_avg2n}. 
Note that the Green's functions given here represent the
first two steps of a Green's function
renormalization procedure based on the averaging method.
\subsection{Thermal fluctuations}
We will now investigate the effect of noise on the parametric oscillator 
\cite{batista2011cooling}.
We start by adding noise to Eq.~(\ref{parOsc}) and obtain
\begin{equation}
    \ddot x=-\omega^2_0 x-\gamma\dot x+F_p\cos(2\omega t)\;x+R(t),
    \label{DuffingNoise}
\end{equation}
where $R(t)$ is a random function that 
satisfies the statistical averages
$\langle R(t)\rangle=0$
and $\langle R(t)R(t')\rangle= 2T\gamma\delta(t-t')$, according to the fluctuation-dissipation theorem \cite{kubo66}. 
The temperature of the heat bath in which the oscillator (or resonator) is
embedded i $T$. Once we integrate these equations of motion we can show how
classical mechanical noise squeezing and heating occur.
We shall now review the analytical method developed in Refs.
\cite{batista2011cooling, batista2011signal} to study the parametrically-driven
oscillator with noise as given by Eq.~(\ref{DuffingNoise}). 

Using the Green's function we obtain the solution $x(t)$ of 
Eq.~(\ref{DuffingNoise}) in the presence of noise $R(t)$
\begin{align}
x(t)&= x_h(t)+\int_{-\infty}^{\infty}dt'\;G(t,t')R(t'),\\
\dot x(t)&=v(t)= v_h(t)+\int_{-\infty}^{\infty}dt'\;\frac{\partial}{\partial t}G(t,t')R(t'),
\end{align}
where $x_h(t)$ is the homogeneous solution, which in the stable zone decays
exponentially with time; since we assume the pump has been turned on for a long
time, $x_h(t)=0$.
By statistically averaging the  fluctuations of $x(t)$
we obtain
\beqna
\langle x^2(t)\rangle &=& \iint_{-\infty}^{\infty}dt'\;dt''G(t,t')G(t,t'')\langle R(t')R(t'')\rangle=2T\gamma\int_{0}^{\infty}d\tau\;G(t,t-\tau)^2,
\label{x2t_avg}\\
\langle v^2(t)\rangle
&=&2T\gamma\int_{-\infty}^{t}dt'\;\left[\frac{\partial}{\partial
t}G(t,t')\right]^2,
\label{v2_avg}
\eeqna
where $\tau=t-t'$. 

 We notice that by varying the pump amplitude $F_p$ and the detuning
$\Omega$, we can create a continuous family of classical thermo-mechanical
squeezed states, generalizing the experimental results of Rugar and Gr\"utter
\cite{rugar91}. 
An estimate of the time average of the statistically averaged thermal
fluctuations, when $|\beta|>|\delta|$, is given by 
\beqna
 \overline{\langle x^2(t)\rangle}&=&
 \frac{2T\gamma}{\omega^2}\int_0^\infty e^{-\gamma\tau}
 \left\{\left[\cosh(\kappa\tau)\sin(\omega\tau)+\frac{\delta}{\kappa}\sinh(\kappa\tau)\cos( \omega\tau)\right]^2
 +\frac{\beta^2}{2\kappa^2}\sinh^2(\kappa\tau)\right\}\,d\tau\nn\\
&=&\frac{2T\gamma}{\omega^2} \left[I_1+I_2+I_3+I_4\right],
\label{x2_avg_dc2}
\eeqna 
where the integrals are given by 
\beqna
I_1&=&\frac{\beta^2}{2\kappa^2}\int_0^\infty e^{-\gamma\tau}\sinh^2(\kappa\tau)
d\tau=
\frac{\beta^2}{\gamma(\gamma^2-4\kappa^2)},\nn\\
I_2&=&\int_0^\infty e^{-\gamma\tau}\cosh^2(\kappa\tau) \sin^2(\omega\tau) d\tau\nn\\
&=&\frac{1}{2}\left\{\frac{1}{2\gamma}-\frac{\gamma}{2(\gamma^2+4\omega^2)}
+\frac{\gamma}{2(\gamma^2-4\kappa^2)}-\frac{1}{4}\mbox{Re}\left[\frac{1}{\gamma-2\kappa-2i\omega}+\frac{1}{\gamma+2\kappa-2i\omega}\right]\right\},\nn\\
I_3&=&\frac{\delta^2}{\kappa^2}\int_0^\infty e^{-\gamma\tau}\sinh^2(\kappa\tau)
\cos^2(\omega\tau) d\tau\nn\\ &=&\frac{1}{\kappa^2}\left\{\frac{\kappa^2}{\gamma(\gamma^2-4\kappa^2)}+\frac{1}{8}\mbox{Re}\left[\frac{1}{\gamma-2\kappa-2i\omega}+\frac{1}{\gamma+2\kappa-2i\omega}\right]-\frac{\gamma}{4(\gamma^2+4\omega^2)}\right\},\nn\\
\label{I_3}
I_4&=&\frac{\delta}{2\kappa}\int_0^\infty
e^{-\gamma\tau}\sinh(2\kappa\tau)\sin(2\omega\tau)
d\tau=\frac{1}{4}\mbox{Im}\left[
\frac{1}{\kappa(\gamma-2\kappa-2i\omega)}-\frac{1}{\kappa(\gamma+2\kappa-2i\omega)}\right].\nn
\eeqna 

An estimate of the statistically averaged thermal fluctuations, when $|\beta|>|\delta|$, is given by

\beqna
\langle x^2(t)\rangle&\approx&\overline{\langle x^2(t)\rangle}+A_{2\omega}\cos(2\omega t)+B_{2\omega}\sin(2\omega t)+A_{4\omega}\cos(4\omega t)
+B_{4\omega}\sin(4\omega t),
\label{x2_avg2}
\eeqna
where
\beqna
A_{2\omega}&=& -\frac{4\beta T\gamma}{\omega^2}(K_1+K_2),\nn\\
B_{2\omega}&=& -\frac{4\beta T\gamma}{\omega^2}(K_3+K_4),\nn
\eeqna
with
\beqna
K_1&=&
\frac{1}{8}\mbox{Im}\left[ \frac{1}{\kappa(\gamma-2\kappa-2i\omega)}-\frac{1}{\kappa(\gamma+2\kappa-2i\omega)}\right],\nn\\
K_2&=&\frac{\delta}{\kappa^2}\left\{\frac{\kappa^2}{\gamma(\gamma^2-4\kappa^2)}-\frac{\gamma}{4(\gamma^2+4\omega^2)}+\frac{1}{8}\mbox{Re}\left[\frac{1}{\gamma-2\kappa-2i\omega}+\frac{1}{\gamma+2\kappa-2i\omega}\right]\right\},\nn\\
K_3&=&\frac{1}{8\kappa}\left[\frac{4\kappa}{(\gamma^2-4\kappa^2)}+\mbox{Re}\left( \frac{1}{\gamma-2\kappa-2i\omega}-\frac{1}{\gamma+2\kappa-2i\omega}\right)
\right],\nn\\
K_4&=&\frac{\delta}{8\kappa^2}\mbox{Im}\left[\frac{1}{\gamma-2\kappa-2i\omega}+\frac{1}{\gamma+2\kappa-2i\omega}-\frac{2}{\gamma-2i\omega}\right]\nn\\
&=&\frac{\delta\omega}{4\kappa^2}\left[\frac{1}{(\gamma+2\kappa)^2+4\omega^2}+\frac{1}{(\gamma-2\kappa)^2+4\omega^2}-\frac{2}{\gamma^2+4\omega^2}\right].\nn
\eeqna
The remaining coefficients of Eq.~(\ref{x2_avg2}) are given by
\beqna
A_{4\omega} &=&\frac{\beta^2 T\gamma}{4\omega^2\kappa^2}\mbox{Re}\left[\frac{1}{\gamma-2\kappa-2i\omega}+\frac{1}{\gamma+2\kappa-2i\omega}-\frac{2}{\gamma-2i\omega}\right],\nn\\
B_{4\omega} &=&\frac{\beta^2 T\gamma}{4\omega\kappa^2}\mbox{Im}\left[\frac{1}{\gamma-2\kappa-2i\omega}+\frac{1}{\gamma+2\kappa-2i\omega}-\frac{2}{\gamma-2i\omega}\right].\nn
\eeqna
%Mencionar comportamento proximo aa zona de instabilidade
Notice that when one gets close to the zone of instability, we obtain a far
simpler expression for the average fluctuations. It is given approximately by

\beq
\langle x^2(t)\rangle \approx \frac{2T}{\omega^2}
\left[ \beta^2+\frac{\gamma^2}{4}+\delta^2
\right]\frac{1}{\gamma^2-4\kappa^2}-\frac{4\beta
T}{\omega^2(\gamma^2-4\kappa^2)}\left[\delta\cos(2\omega
t)+\frac{\gamma}{2}\sin(2\omega t)\right].
\label{x2_approx}
\eeq
It is easy to verify that the minimum of the position fluctuation is given by 
\[
\sigma_{\mbox{min}}^2\equiv\langle x^2(t)\rangle_{\mbox{min}} \approx
\frac{T}{2\omega^2}\frac{\sqrt{\delta^2+\gamma^2/4}-|\beta|}{\sqrt{\delta^2+\gamma^2/4}+|\beta|}
\]
and the maximum is given by
\[
\sigma_{\mbox{max}}^2\equiv{\langle x^2(t)\rangle}_{\mbox{max}} \approx
\frac{T}{2\omega^2}\frac{\sqrt{\delta^2+\gamma^2/4}+|\beta|}{\sqrt{\delta^2+\gamma^2/4}-|\beta|},
\]
such that 
\beq
{\sigma}_{\mbox{min}} {\sigma}_{\mbox{max}}\approx
\frac{T}{2\omega^2}.
\label{squeezing}
\eeq
This is a verification that the classical phenomenon of thermal squeezing indeed
occurs near the transition line to instability.
These approximate expressions for squeezing are not valid with detuning
($\delta \neq 0$) when $\beta=0$, although they correctly predict no squeezing
in such situation and the equipartition theorem is correct to $O(\epsilon)$ if
$\Omega=O(\epsilon)$.

From Eqs. \eqref{green_avg2n} and \eqref{x2t_avg}  we obtain the dc
contribution to fluctuations in the second-order approximation
\beqna
\overline{\langle x^2(t)\rangle} &\approx&
\frac{2\gamma T}{\omega^2}\int_{0}^{\infty}d\tau\;e^{-\gamma\tau}\left\{
\left[\cosh(\xi \tau)\sin(\omega\tau) +c\frac{\sinh(\xi \tau)}{\xi}
\cos(\omega\tau)\right]^2+\beta'^2\frac{\sinh^2(\xi\tau)}{2\xi^2}\right\}\nn\\
&=&\frac{2T\gamma}{\omega^2} \left[\tilde I_1+\tilde I_2+\tilde
I_3+\tilde I_4\right],
\label{x2_avg2_approx}
\eeqna 
where the integrals are given by
\beqna
\tilde I_1&=& \frac{\beta'^2}{\gamma(\gamma^2-4\xi^2)},\nn\\
\tilde I_2&=&\frac{1}{2}\left\{\frac{1}{2\gamma}-\frac{\gamma}{2(\gamma^2+4\omega^2)}
+\frac{\gamma}{2(\gamma^2-4\xi^2)}-\frac{1}{4}\mbox{Re}\left[\frac{1}{\gamma-2\xi-2i\omega}+\frac{1}{\gamma+2\xi-2i\omega}\right]\right\},\nn\\
\tilde I_3&=&\frac{1}{\xi^2}\left\{\frac{\xi^2}{\gamma(\gamma^2-4\xi^2)}+\frac{1}{8}\mbox{Re}\left[\frac{1}{\gamma-2\xi-2i\omega}+\frac{1}{\gamma+2\xi-2i\omega}\right]-\frac{\gamma}{4(\gamma^2+4\omega^2)}\right\},\nn\\
\label{I2_3}
\tilde I_4&=&\frac{1}{4}\mbox{Im}\left[
\frac{1}{\xi(\gamma-2\xi-2i\omega)}-\frac{1}{\xi(\gamma+2\xi-2i\omega)}\right].\nn
\eeqna 
All the other first-order approximation results given at
Eqs.~(\ref{x2_avg2}-\ref{squeezing}) are the same in second-order
approximation, except for the parameter replacements given at
the end of subsubsection (\ref{2nd-orderGF}).
It is noteworthy to mention that the squeezing condition given at first-order
approximation in Eq.~\eqref{squeezing} is still valid in the second-order
approximation.
This implies that a renormalization procedure based on the averaging method
is possible for the Green's function of the parametric oscillator with
parameters set near the onset of the first instability zone.
\subsection{Energy balance}
It is important to verify that in the stable zone of
the parametric oscillator, on average, the power input from the external noise
and from the internal pump is balanced by the dissipated power. 
 
 The instantaneous input power due to the additive noise is given by
\beq
P_{noise}(t)=\dot x(t)R(t)=
v_h(t)R(t)+\int_{-\infty}^{\infty}dt'\;\frac{\partial}{\partial
t}G(t,t')R(t')R(t).
\eeq
Hence, we obtain the statistically averaged  noise input power
\beq
\langle P_{noise}(t)\rangle=
\int_{-\infty}^{t}dt'\;\frac{\partial}{\partial
t}G(t,t')\langle R(t')R(t)\rangle=\gamma T \left[\frac{\partial}{\partial
t}G(t,t')\right]_{t'=t^-}=\gamma T.
\label{noise}
\eeq
The statistically-averaged dissipated power is given by
\beqna
\langle P_{diss}(t)\rangle &=& -\gamma\langle \dot x(t)^2\rangle=
-\gamma \int_{-\infty}^{t}dt'\int_{-\infty}^{t}dt''\;
\frac{\partial}{\partial t}G(t,t')\frac{\partial}{\partial
t}G(t,t'')\langle R(t')R(t'')\rangle\nn\\
&=&-2\gamma^2 T
\int_{-\infty}^{t}dt'\;\left[\frac{\partial}{\partial
t}G(t,t')\right]^2.
\label{dissPower}
\eeqna
The statistically-averaged pump power is given by
\beqna
\langle P_{pump}(t)\rangle &=& F_p\cos(2\omega t)\langle x(t)\dot
x(t)\rangle\nn\\
&=& F_p\cos(2\omega
t)\int_{-\infty}^{t}dt'\;\int_{-\infty}^{t}dt''\;G(t,t')\frac{\partial}{\partial
t}G(t,t'') \langle R(t')R(t'')\rangle\nn\\
&=&2\gamma T F_p\cos(2\omega
t)\int_{-\infty}^{t}dt'\;G(t,t')\frac{\partial}{\partial t}G(t,t')\nn\\
&=&
\gamma T F_p\cos(2\omega
t)\frac{\partial}{\partial t}\int_{-\infty}^{t}dt'\;G(t,t')^2\nn\\
&=&
\frac{F_p}{2}\cos(2\omega
t)\frac{\partial}{\partial t}\langle x(t)^2\rangle.
\eeqna
From Eq.~(\ref{x2_avg2}) we obtain approximately the time-averaged 
statistically-averaged pump power.
 It is given by
\beq
\overline{ \langle P_{pump}(t)\rangle}\approx\frac{F_p\omega}{2}B_{2\omega}.
\eeq
Near the threshold to instability in first-order approximation it becomes
\[
\overline{ \langle P_{pump}(t)\rangle}
\approx \frac{F_p^2
T\gamma}{4\omega^2(\gamma^2-4\kappa^2)},
\]
while on second-order approximation it is
\[
\overline{ \langle P_{pump}(t)\rangle}\approx
\frac{F_p^2
T\gamma}{4\omega^2(\gamma^2-4\xi^2)}\sqrt{(1-\delta/\omega)^2+\gamma^2/(4\omega^2)},
\]

In the stable zone of the parametric oscillator, when the stationary point is reached, 
one gets the energy balance on average, that is
\beq
\overline{\langle P_{noise}(t)\rangle}+\overline{\langle
P_{diss}(t)\rangle}+\overline{\langle P_{pump}(t)\rangle}=0.
\label{energyBalance}
\eeq
One can use the above expression to obtain the time average of the velocity
fluctuations
\[
\overline{\langle \dot x(t)^2\rangle}= \left(T+\overline{\langle
P_{pump}(t)\rangle}/\gamma\right)=\left(T+\frac{F_p\omega}{2\gamma}B_{2\omega}\right).
\]
By comparison of the above equation with Eq.~\eqref{x2_avg2_approx} one notices
that the equipartion of energy breaks down once pumping is on.
\section{Results and Discussion}
\label{numerics}
\begin{figure}[!h]
\includegraphics[scale=1.0]{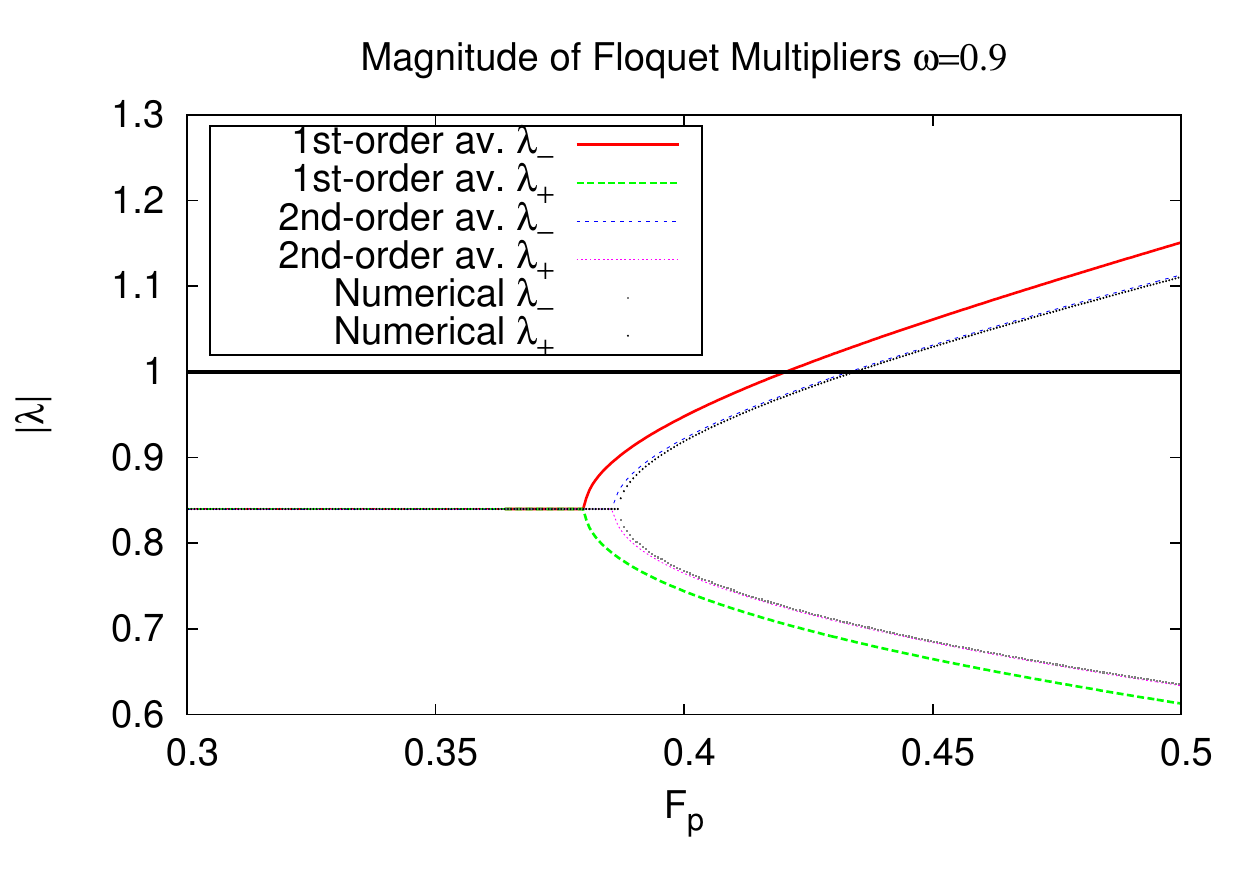}
\caption{(color online) The magnitude of the Floquet multipliers at $\omega=0.9,
\gamma=0.1$. The transition to instability of the dynamics of the parametric oscillator occurs when one of the Floquet
multipliers becomes bigger than 1. The second-order averaging prediction
[Eq.~(\ref{2ndOrderFM})] of the FMs reproduces very accurately the numerical
results. The branching off of the FMs occurs when they become real.
}
\label{fmw0.9}
\end{figure}

\begin{figure}[!h]
\includegraphics[scale=1.0]{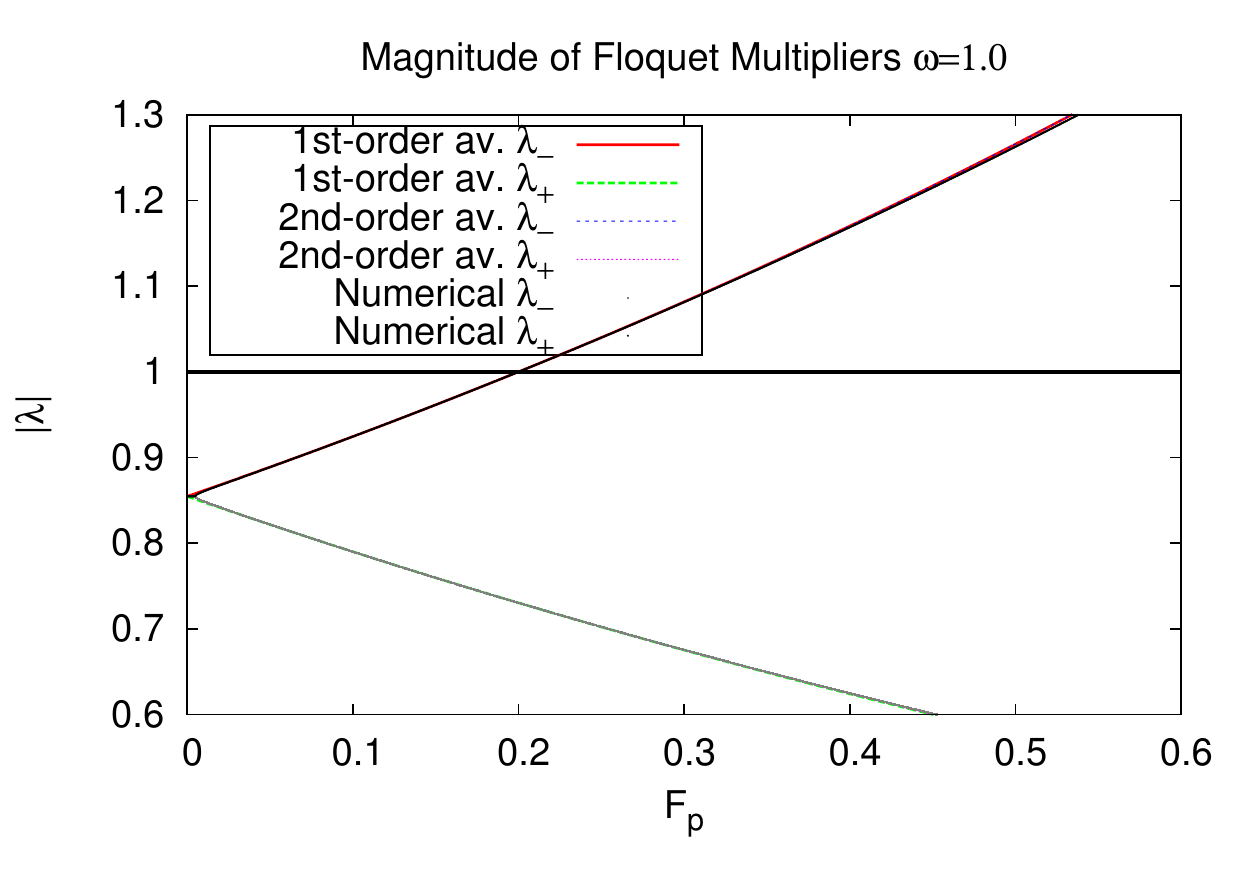}
\caption{(color online) The magnitude of the Floquet multipliers at $\omega=1.0,
\gamma=0.1$. From the onset of the parametric drive the FMs are real. The second-order averaging prediction
[Eq.~(\ref{2ndOrderFM})] of the FMs reproduces very accurately the numerical
results.}
\label{fmw1}
\end{figure}

\begin{figure}[!h]
\includegraphics[scale=1.0]{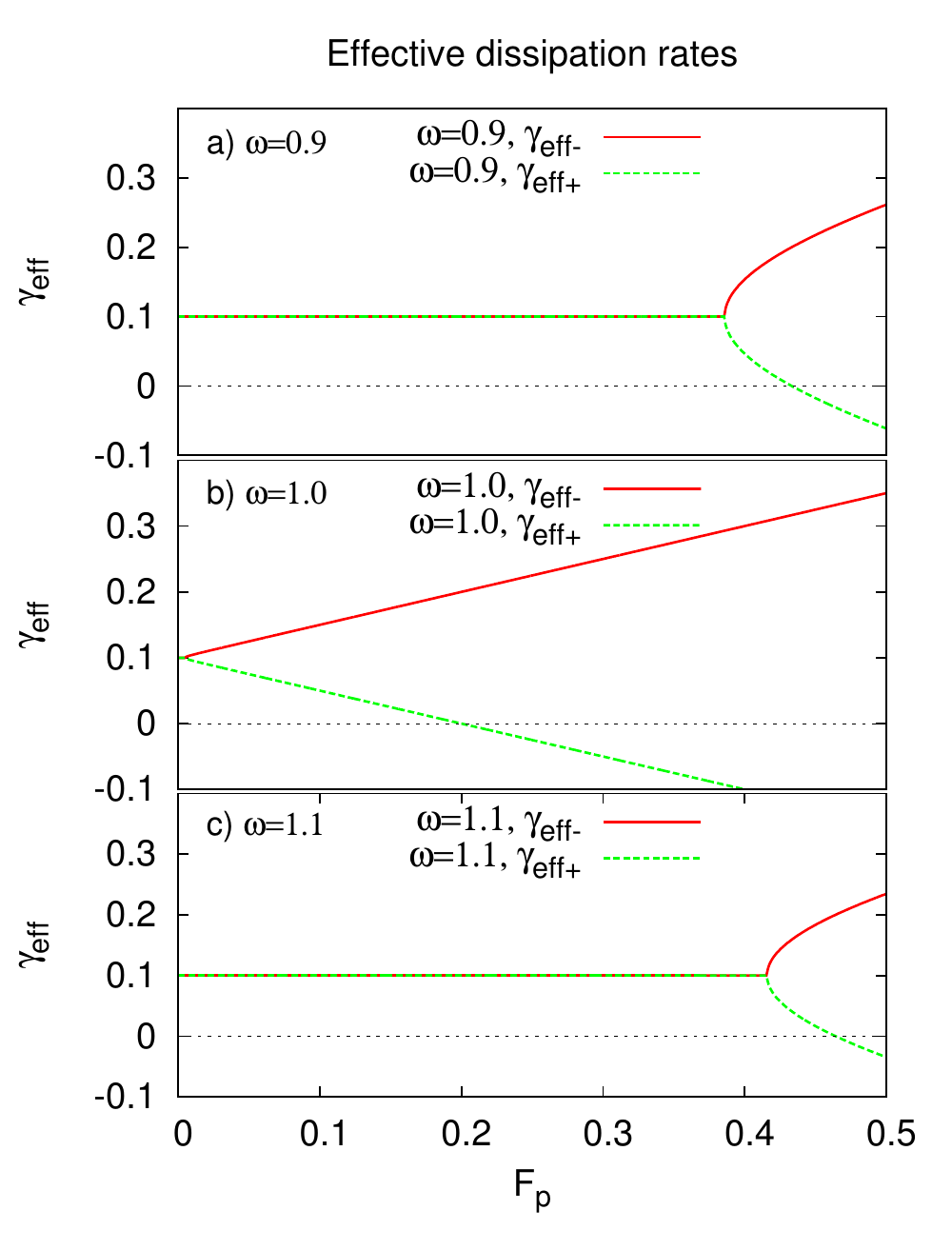}
\caption{(color online) Effective dissipation rates based on
second-order averaging approximation obtained from Eq.~(\ref{2ndOrderFM}).
Heating and squeezing mostly occur when the floquet multipliers become real and
branch off in two different values.
Consequently, there is an effective dissipation rate associated with each
floquet multiplier. An effective dissipation rate $\gamma_{eff}<\gamma$ cause
heating. The different values of $\gamma_{eff}$, one smaller
than $\gamma$ and the other larger than $\gamma$, result in squeezing.
$\gamma=0.1$}
\label{e-diss}
\end{figure}

\begin{figure}[!h]
\includegraphics[scale=1.0]{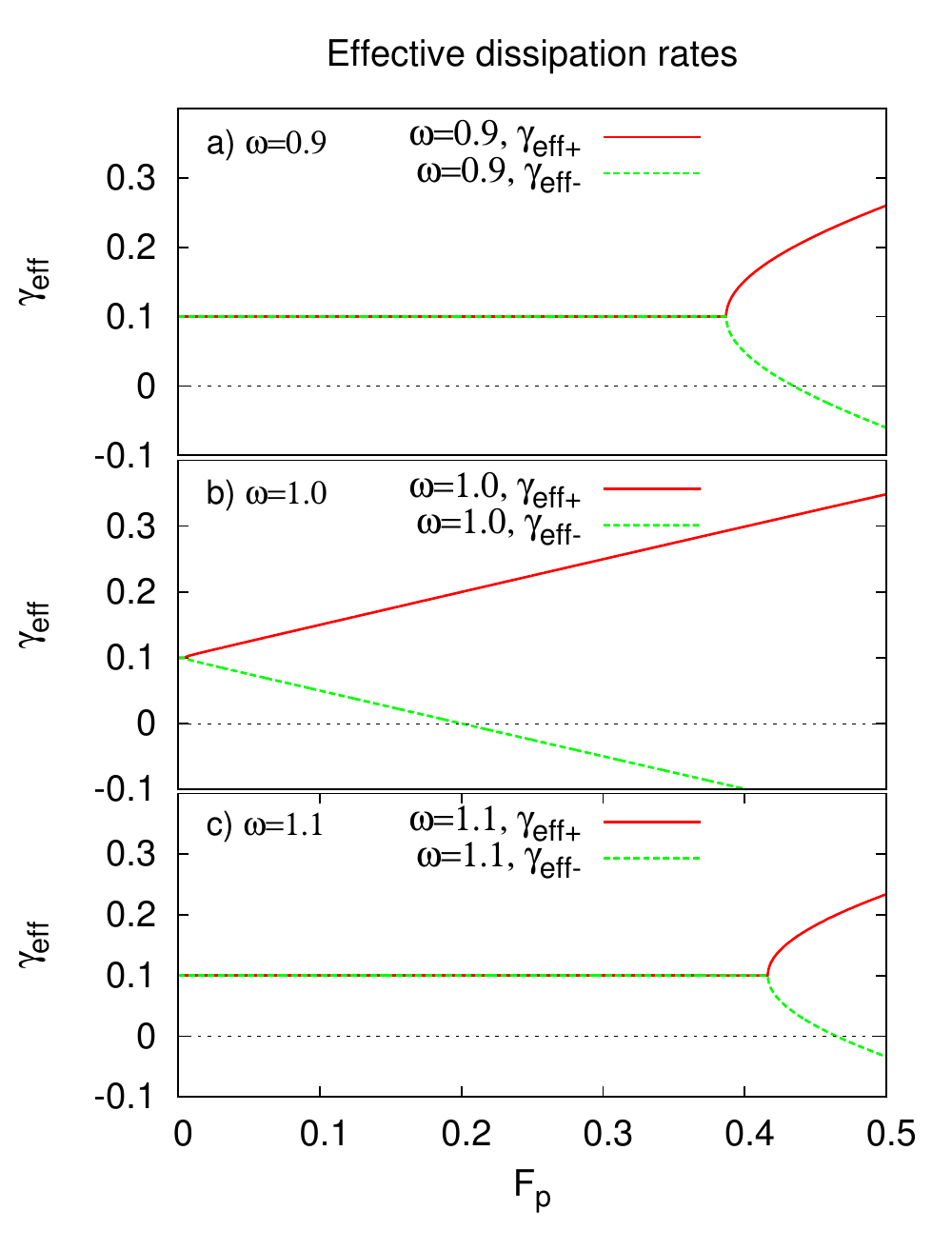}
\caption{(color online) Numerical effective dissipation rates. Same parameters
as in Fig.~(\ref{e-diss}).}
\label{e-dissN}
\end{figure}

In Figs.~(\ref{fmw0.9}, \ref{fmw1}, \ref{fmw1.1}) we show the dependance of the
magnitude of the Floquet multipliers (FMs) on the pump amplitude $F_p$. 
When the FMs given by Eq.~(\ref{2ndOrderFM}) become real, they branch off in two
magnitudes. 
When this occurs one of the FMs ($\lambda_-$) becomes larger as $F_p$ is
increased, eventually becoming larger than one, when the system given by
Eq.~(\ref{parOsc}) becomes unstable, while the other FM ($\lambda_+$)
becomes smaller. From Eq.~(\ref{2ndOrderFM}) this implies into two effective dissipation rates, 
along the stable manifolds of the Poincaré first-return map (with period
$\pi/\omega$, i.e. $\Phi(\pi/\omega)$). 
This phenomenon causes both heating and squeezing of the parametric oscillator
with added noise, since the average power input due to thermal noise remains constant
as is shown in Eq.~(\ref{noise}) and one effective dissipation rate is
decreased.
It also causes quadrature thermal squeezing since in one direction,
the $\lambda_-$-stable manifold of $\Phi (\pi/\omega)$ in the phase space of $u$
and $v$, there is less effective dissipation and, consequently, more
fluctuations, while in another direction, the $\lambda_+$-stable manifold of
$\Phi (\pi/\omega)$ in the phase space of $u$ and $v$, there is more effective
dissipation and , consequently, less fluctuations.
This inbalance in the effective dissipation rates, we claim, is the main cause
of thermal squeezing in the parametric oscillator with added noise.
The dependance on pump amplitude of the effective dissipations can be seen on
Fig.~\ref{e-diss} (second-order averaging result) and on
Fig.~\ref{e-dissN} (Floquet theory numerical result).

\begin{figure}[!h]
    \includegraphics[scale=1.0]{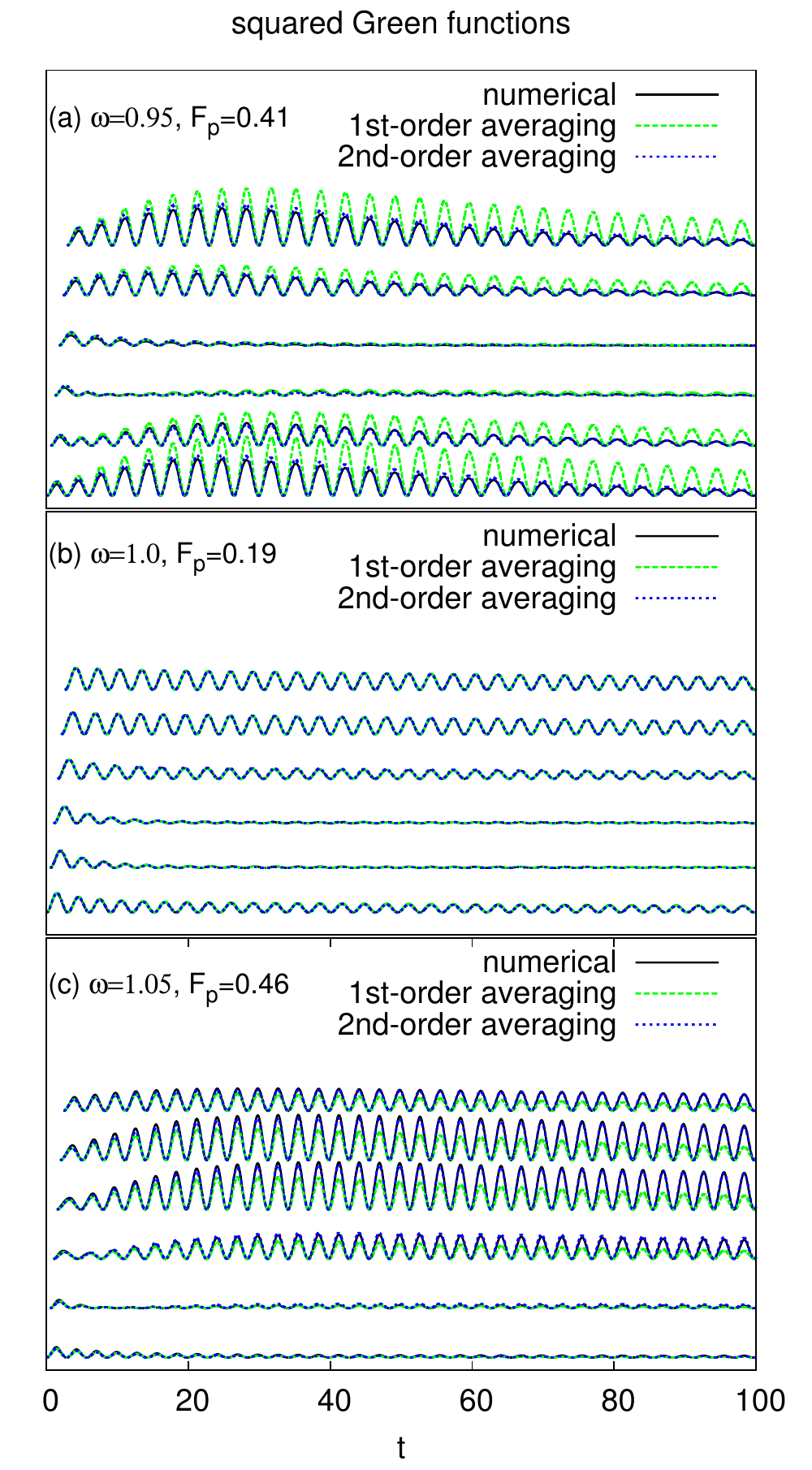}
    \caption{
    (color online) In the frames above we show several squared Green's functions
    with equally-spaced in time initial conditions in one given period of the
    parametric driving. They are vertically spaced for clarity, since all
    asymptotes are zero. In each frame we have a comparison between numerical
    results given by the numerical integration of Eq.~(\ref{green_eq}) and the
    analytical approximate results given by Eqs.~(\ref{green_avg}) or
    (\ref{green_avg2}). We have a) $\omega=0.9$, b)
    $\omega=1.0$, and c) $\omega=1.1$. The initial values of
    the Green's functions are $G(t,t')=0$ and $\frac{\partial}{\partial
    t}G(t,t')=1.0$ when $t=t'+0^+$. The pump amplitudes are indicated in the
    figure frames.}
    \label{greenfs}
\end{figure}

In Fig.~\ref{greenfs} we show several squared Green's functions with initial
conditions spread out evenly in time during one period of the pump
($\pi/\omega$). They are vertically spaced only for clarity, since all of their assymptotes are
zero.
The Green's functions are shown squared because that is the way they contribute
to the thermal fluctuations in Eq.~(\ref{x2t_avg}). 
One notices that the second-order approximation Green's functions yield a much
better approximation to the numerical Green's functions than the first-order
approximation.
This is specially evidenced the closer one gets to first transition line to
instability, a consequence of the fact that the second-order analytical
expression for the transition line given in Eq.~(\ref{2ndOrderTransLine}) is
a better approximation than the first-order expression in Eq.~(\ref{eq_zones}).

\begin{figure}[!h]
    \includegraphics[scale=1.0]{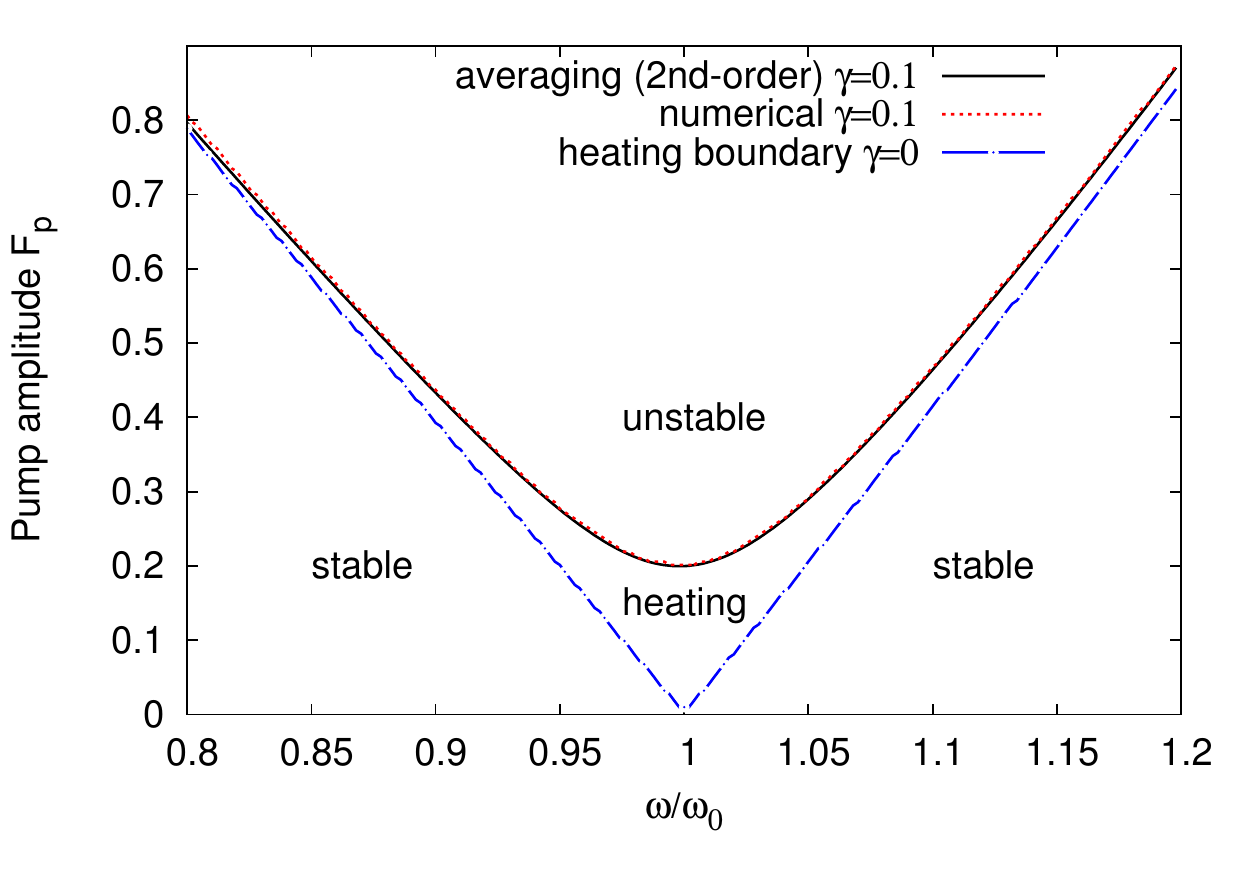}
\caption{ (color online) 
Comparison between numerical and averaging method predictions for the boundary
of the first instability zone of the damped parametrically-driven oscillator of
Eq.~\eqref{parOsc}. In the region above the thick solid black line lies the
unstable zone obtained by numerical computation, the red dashed line  is the second-order averaging 
prediction [Eq.~\eqref{2ndOrderTransLine}] for the transition line. The
numerical results are obtained by numerically calculating the corresponding
Floquet multipliers. 
The transition line to parametric instability is defined when at least one of
them has modulus equal to 1. The heating zone occurs when the Floquet exponents
given in Eq.~\eqref{2ndOrderFM} become real. The heating boundary (blue long
dashed-dotted line) is equivalent to the transition line to parametric
instability when $\gamma=0$. }
    \label{zones}
\end{figure}

\begin{figure}[!h]
\includegraphics[scale=1.0]{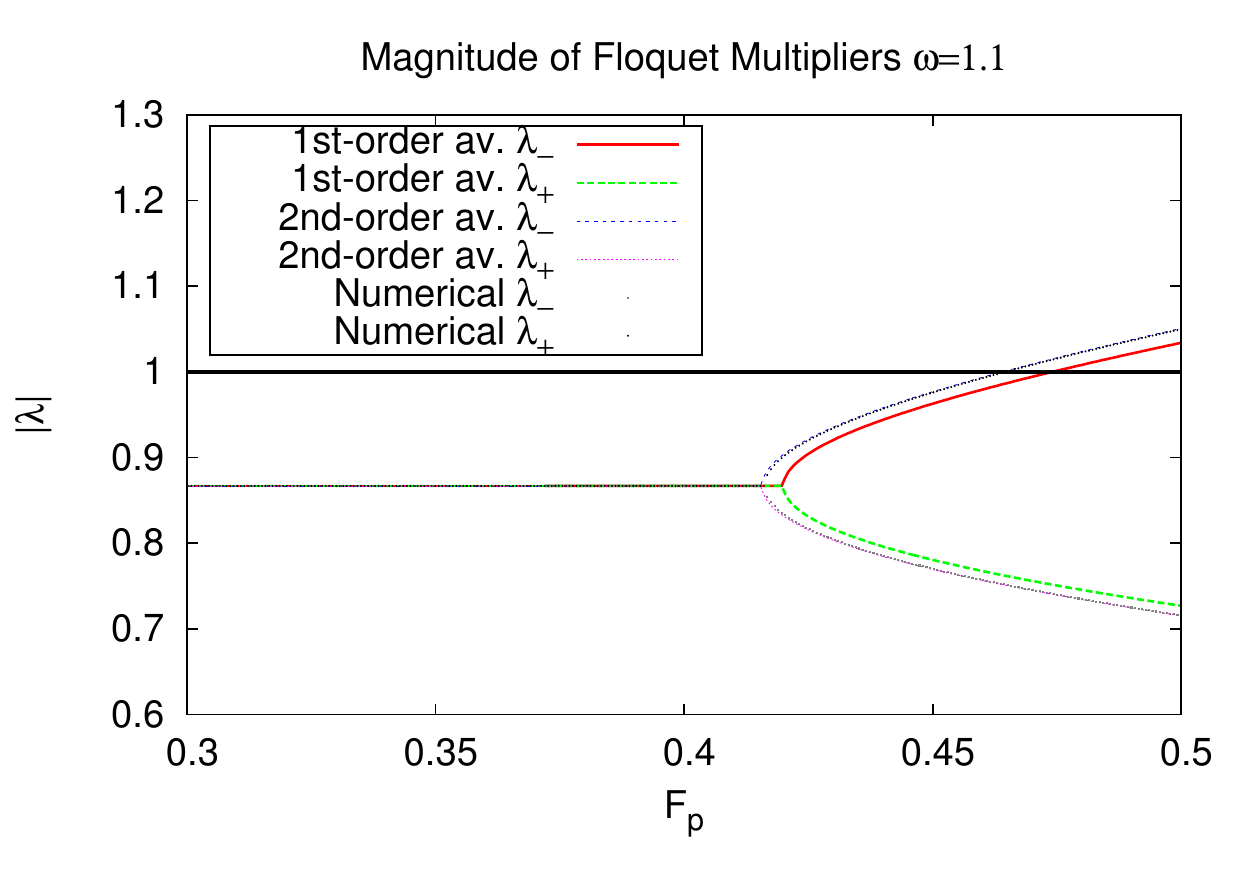}
\caption{(color online) The magnitude of the Floquet multipliers at $\omega=1.1,
\gamma=0.1$. The second-order averaging prediction
[Eq.~(\ref{2ndOrderFM})] of the FMs reproduces very accurately the numerical
results. Off resonance one notices that the pump amplitude has to be higher before the
branching off of the Poincaré map eigenvalues (FMs) in agreement with the
results of Fig.~(\ref{zones}). }
\label{fmw1.1}
\end{figure}

\begin{figure}[!h]
\includegraphics[scale=1.0]{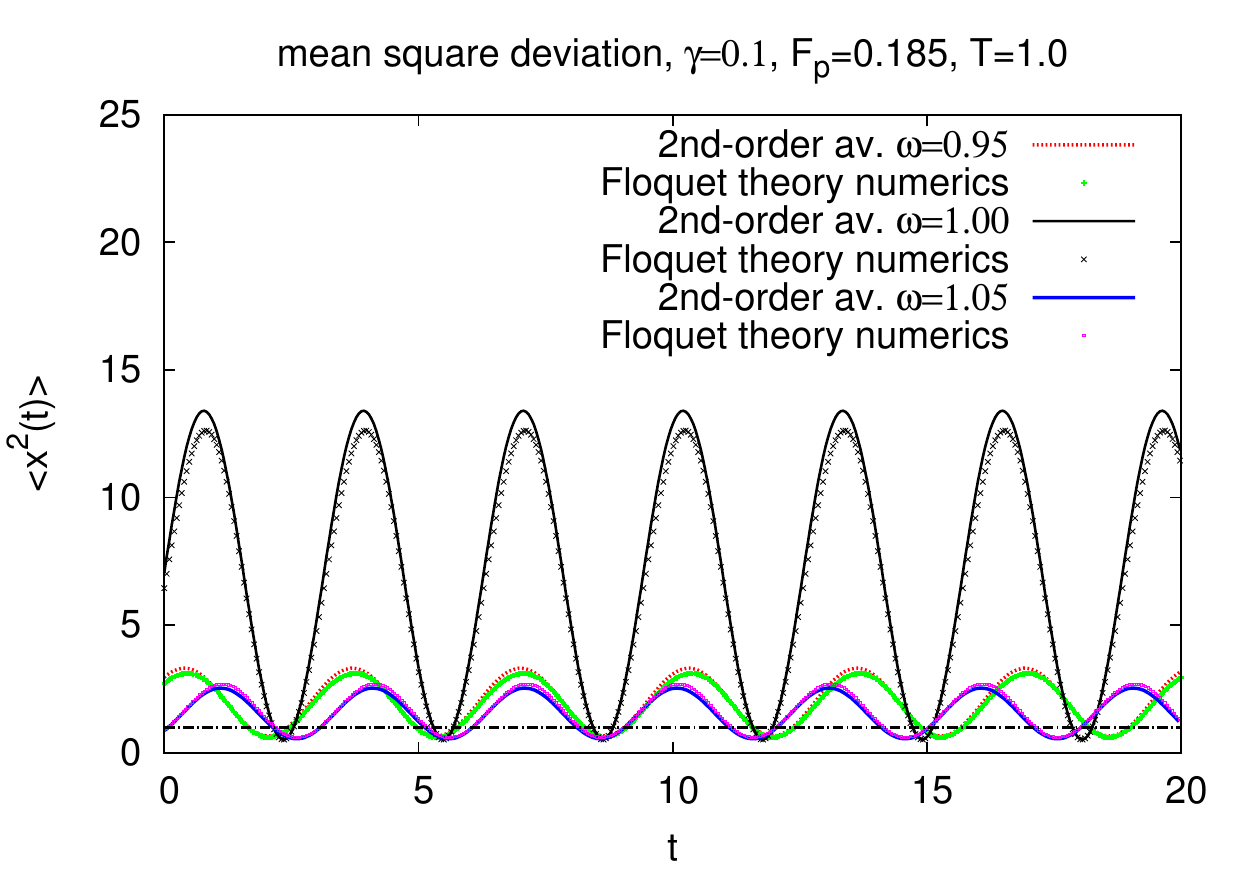}
\caption{Mean-square deviation $\langle x^2(t)\rangle$ time evolution.
Comparison between Floquet theory numerical results and second-order averaging
results given by Eq.~(\ref{x2t_avg}) with the Green's function given by
Eq.~(\ref{green_avg2}). 
The dot-long dashed line indicates the temperature of the external heat bath.
All pump amplitudes are the same $F_p=0.185$.
This indicates that the close one is to resonance the larger the squeezing
amplitude and average dynamical temperature. }
\label{x2_t}
\end{figure}

\begin{figure}[!h]
\includegraphics[scale=1.0]{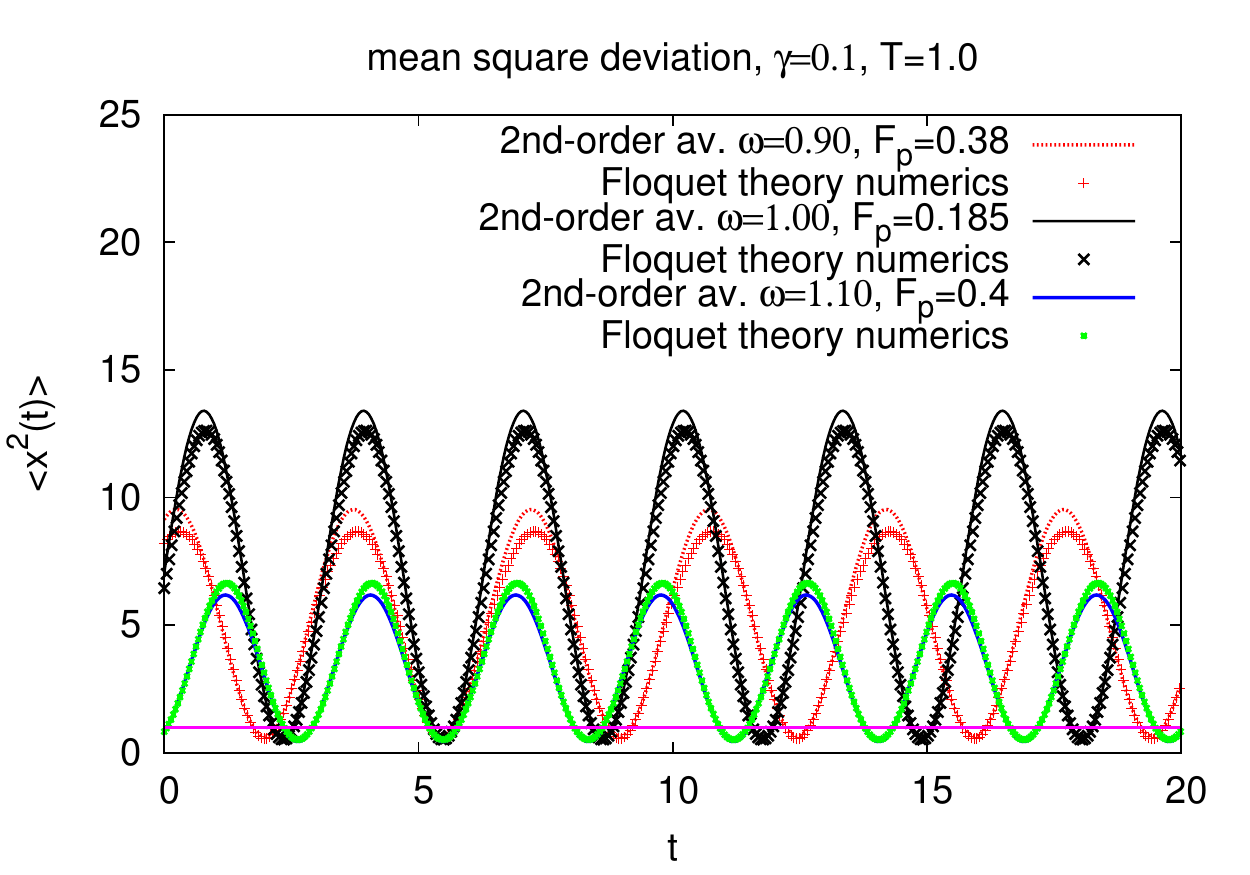}
\caption{Mean-square deviation $\langle x^2(t)\rangle$ time evolution.
Comparison between Floquet theory numerical results and second-order averaging
results given by Eq.~(\ref{x2t_avg}) with the Green's function given by
Eq.~(\ref{green_avg2}). 
The dot-long dashed line indicates the temperature of the external heat bath.
The values of the pump amplitude were chosen to be close to the transition line
given by Eq.~(\ref{2ndOrderTransLine})}.
\label{x2_floquet}
\end{figure}

In Fig.~\ref{x2_t} we show a comparison between time series of the
fluctuation $\langle x^2(t)\rangle$ given by Eq. (\ref{x2t_avg}) in which the Green's functions are
given either by the second-order approximation expression from
Eq.~(\ref{green_avg2}) or by the numerical Floquet theory Green's functions. 
We use several different Green's functions with negative detuning
($\omega=0.95$), in resonance ($\omega=1.0$) and positive detuning
($\omega=1.05$) all with the same pump amplitude $F_p=0.185$.
We observe that the numerical and approximate Green's functions are very
similar and that the squeezing amplitude and heating (proportional to the time
average of $\langle x^2(t)\rangle$) are very dependant on detuning from
resonance. From Fig. \ref{zones} one sees that the off resonance results are
slightly below the threshold of the strong heating and squeezing zone.
In Fig.~\ref{x2_floquet} we show again several time series of the fluctuation
$\langle x^2(t)\rangle$, but this time the pump amplitudes are chosen such that
the parameters are inside the heating zone (as given in Fig.~\ref{zones}) and
close to the transition line to instability. One sees then  considerably higher dynamical temperatures and squeezing
amplitudes for the detuned fluctuations than in the previous figure.

\begin{figure}[!h]
    \includegraphics[scale=1.0]{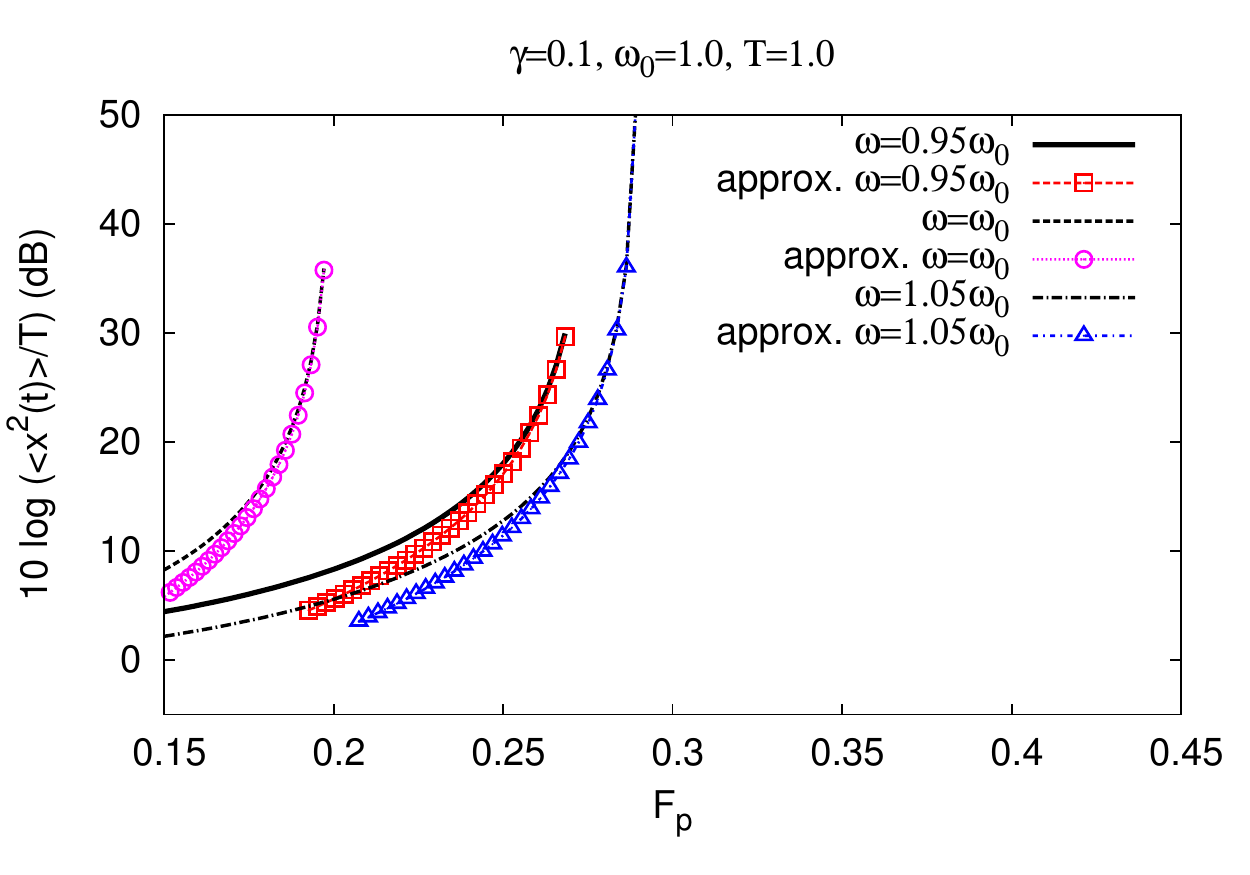}
    \caption{
    (color online) Log plot of the dc component of the mean square displacement
    in second-order approximation, $\overline{\langle x^2(t)\rangle}$, as given
    by Eqs.~(\ref{x2_avg2_approx}). In the parametric oscillator
    with thermal noise the dynamic temperature of the oscillator grows
    monotonically until it diverges at the transition line between stable and
    unstable zones. The simplified approximating curves are given by
    Eq.~(\ref{x2_approx}). }
    \label{heating}
\end{figure}

In Fig.~\ref{heating} we show a logarithmic plot of the dc component of
the mean-square displacement over the heat bath temperature.
Most of the heating occurs inside the heating zone in which the Floquet
multipliers are real, each one with a different amplitude, one that increases 
and the other that decreases as the pump amplitude is increased.

\begin{figure}[!h]
    \includegraphics[scale=1.0]{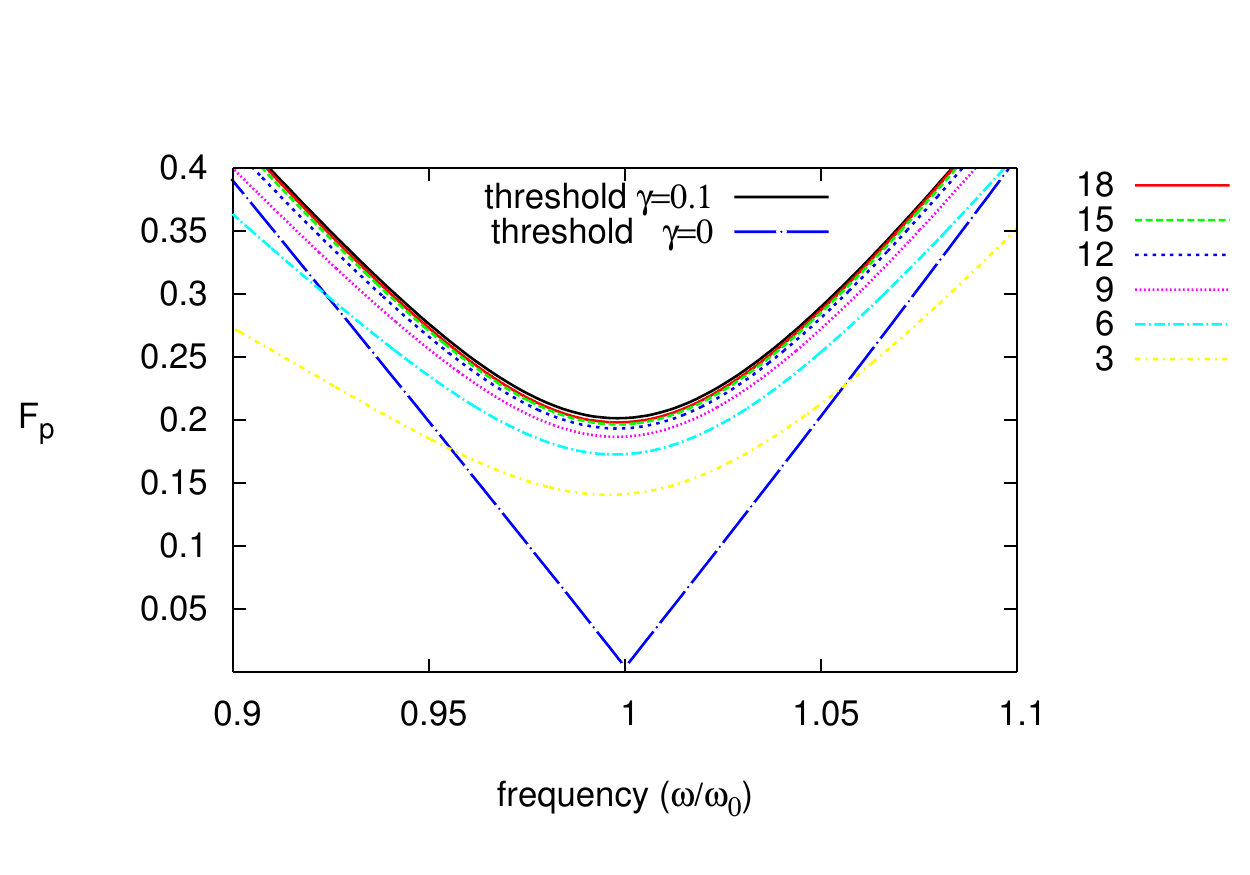}
    \caption{
    (color online) Countour log plot of dynamic temperatures (the dc component
    of the statistical average of the square displacement in second-order
    approximation, $\overline{\langle x^2(t)\rangle}$ , as given by Eqs.~\eqref{x2_avg2_approx}) over
    heat bath temperature. Most of the heating occurs inside the heating zone
    as predicted by our hypothesis. Each level curve is a dynamic isothermal. 
    The levels are given in decibels given by the expression $10
    \log(\overline{\langle x^2(t)\rangle}/T)$, hence each increment of
    approximately 3 dB increases the dynamic temperature by a factor of 2. }
    \label{tempcont}
\end{figure}

\begin{figure}[!h]
    \includegraphics[scale=1.0]{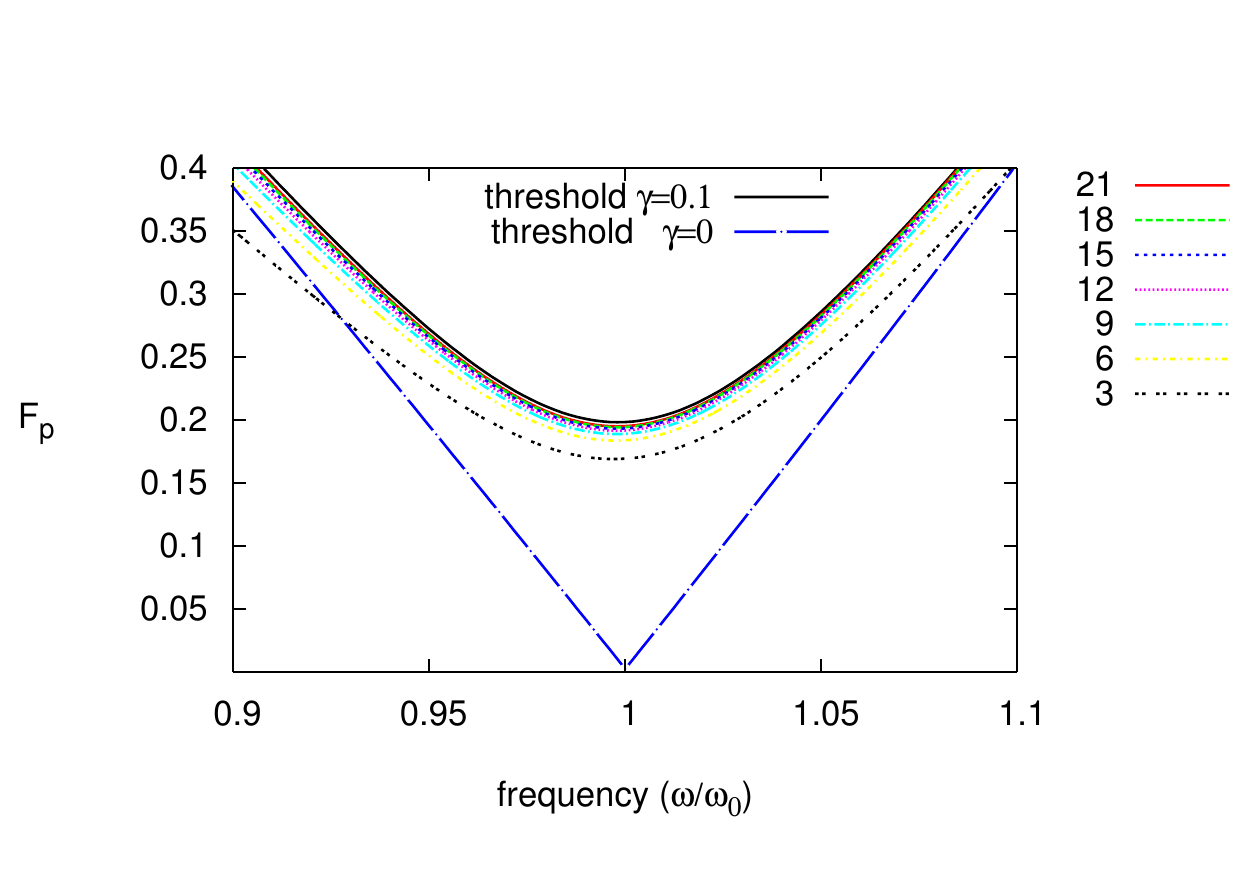}
    \caption{
    (color online) Countour log plot of the squeezing amplitude (the dc
    component of the statistical average of the square displacement in
    second-order approximation, $\sqrt{A_{2\omega}^2+B_{2\omega}^2}$ , as given
    by Eqs.~\eqref{x2_avg2}) (with the appropriate second-order corrections) over
    heat bath temperature. Most of the squeezing occurs inside the heating zone as predicted by our hypothesis.
    Each level curve has constant squeezing amplitude. The levels are given in
    decibels given by the expression $10
    \log(\sqrt{A_{2\omega}^2+B_{2\omega}^2}/T)$. }
    \label{sqzAmp}
\end{figure}

In Fig.~\ref{tempcont} we show level sets in decibels of the dc
component of the mean-square displacement  over the heat bath
temperature in second-order approximation. One sees that most of the heating
occurs inside the heating zone in which the Floquet multipliers are real.
In Fig.~\ref{sqzAmp} we show level sets of the squeezing amplitude of
the mean-square displacement over the heat bath temperature in second-order
approximation. This attests that most of the squeezing also occurs inside the
heating zone as claimed before.

\section{Conclusions}
\label{conclusions}
% What we did here
Here, we studied  a parametrically-driven
oscillator with added noise with the objective of understanding what
causes heating and thermal squeezing in the stable zone near the
transition line of the first parametric instability. 
% Statement of our objectives
We improved on our previous work \cite{batista2011cooling}
and obtained a more accurate expression for the Green's functions
and, consequently, obtained more precise estimates of the amount of
heating and quadrature noise squeezing  in the parametrically-driven oscillator.
%
% Description of our theoretical model+achievements
% How we did it
Furthermore, we used an approximate Floquet theory,
based on first and second-order averaging approximations, to explain why 
heating and thermal squeezing occur in the parametrically-driven oscillator
with added noise investigated here. 
These phenomena are related to the onset of real-valued Floquet multipliers
(FMs) with different magnitudes. It was shown that as one FM grows while one gets closer
 (in parameter space) to the first transition line to instability the other FM
decreases. As a consequence, one gets two different effective dissipation rates,
while at the same time the input power due to noise remains constant as the pump
amplitude is increased. 
We showed that these effects account for most thermal squeezing and
heating observed. 
We showed as well that the second-order Green's function of the parametric
oscillator has the same form as the first-order Green's function, which implies
that with a simple parameter change given in the text all first-order results
for the amount of heating and thermal squeezing also apply, with increased
accuracy, to second-order approximation.
This indicates that a Green's function renormalization is very well feasible
if one goes to higher orders of approximation in the averaging method.

% Outlook
The one-degree of freedom model studied here may be applied for instance to the
dynamics of the fundamental mode of an axially loaded doubly-clamped beam
resonator. The present model could also be applied 
to the linear response of ac driven nonlinear oscillators,
such as transversally-loaded beam resonators, to noise.
We note further that this model can be applied as well
to the investigation of the dynamics of ions in quadrupole RF ion guides or
traps \cite{paul90} in the presence of thermal noise. 
Most importantly, the
notion that we can have two different effective dissipation rates, one for each stable Floquet multiplier,  
could be used to create control schemes similar to those of the OGY method
\cite{ott1990controlling} to reduce the effects of noise even further in
parametric amplifiers and in RF ion traps.

%\bibliography{/home/aab/research/artigos/gen2}%>> bibliography data

\begin{thebibliography}{10}%
\makeatletter
\providecommand \@ifxundefined [1]{%
 \ifx #1\undefined \expandafter \@firstoftwo
 \else \expandafter \@secondoftwo
\fi
}%
\providecommand \@ifnum [1]{%
 \ifnum #1\expandafter \@firstoftwo
 \else \expandafter \@secondoftwo
\fi
}%
\providecommand \enquote [1]{``#1''}%
\providecommand \bibnamefont  [1]{#1}%
\providecommand \bibfnamefont [1]{#1}%
\providecommand \citenamefont [1]{#1}%
\providecommand\href[0]{\@sanitize\@href}%
\providecommand\@href[1]{\endgroup\@@startlink{#1}\endgroup\@@href}%
\providecommand\@@href[1]{#1\@@endlink}%
\providecommand \@sanitize [0]{\begingroup\catcode`\&12\catcode`\#12\relax}%
\@ifxundefined \pdfoutput {\@firstoftwo}{%
 \@ifnum{\z@=\pdfoutput}{\@firstoftwo}{\@secondoftwo}%
}{%
 \providecommand\@@startlink[1]{\leavevmode}%
 \providecommand\@@endlink[0]{}%
}{%
 \providecommand\@@startlink[1]{%
  \leavevmode
  \pdfstartlink
   attr{/Border[0 0 1 ]/H/I/C[0 1 1]}%
   user{/Subtype/Link/A<</Type/Action/S/URI/URI(#1)>>}%
  \relax
 }%
 \providecommand\@@endlink[0]{\pdfendlink}%
}%
\providecommand \url  [0]{\begingroup\@sanitize \@url }%
\providecommand \@url [1]{\endgroup\@href {#1}{\urlprefix}}%
\providecommand \urlprefix [0]{URL }%
\providecommand \Eprint[0]{\href }%
\@ifxundefined \urlstyle {%
  \providecommand \doi [1]{doi:\discretionary{}{}{}#1}%
}{%
  \providecommand \doi [0]{doi:\discretionary{}{}{}\begingroup
  \urlstyle{rm}\Url }%
}%
\providecommand \doibase [0]{http://dx.doi.org/}%
\providecommand \Doi[1]{\href{\doibase#1}}%
\providecommand \bibAnnote [3]{%
  \BibitemShut{#1}%
  \begin{quotation}\noindent
    \textsc{Key:}\ #2\\\textsc{Annotation:}\ #3%
  \end{quotation}%
}%
\providecommand \bibAnnoteFile [2]{%
  \IfFileExists{#2}{\bibAnnote {#1} {#2} {\input{#2}}}{}%
}%
\providecommand \typeout [0]{\immediate \write \m@ne }%
\providecommand \selectlanguage [0]{\@gobble}%
\providecommand \bibinfo [0]{\@secondoftwo}%
\providecommand \bibfield [0]{\@secondoftwo}%
\providecommand \translation [1]{[#1]}%
\providecommand \BibitemOpen[0]{}%
\providecommand \bibitemStop [0]{}%
\providecommand \bibitemNoStop [0]{.\EOS\space}%
\providecommand \EOS [0]{\spacefactor3000\relax}%
\providecommand \BibitemShut [1]{\csname bibitem#1\endcsname}%
%</preamble>
\bibitem{faraday1831}%
  \BibitemOpen
  \bibfield{author}{%
  \bibinfo {author} {\bibfnamefont{M.}~\bibnamefont{Faraday}},\ }%
  \bibfield{journal}{%
  \bibinfo {journal} {Philos. Trans. R. Soc. London}\ }%
  \textbf{\bibinfo {volume} {121}},\ \bibinfo {pages} {319} (\bibinfo {year}
  {1831})%
  \bibAnnoteFile{NoStop}{faraday1831}%
\bibitem{ruby1996}%
  \BibitemOpen
  \bibfield{author}{%
  \bibinfo {author} {\bibfnamefont{L.}~\bibnamefont{Ruby}},\ }%
  \bibfield{journal}{%
  \bibinfo {journal} {Am. J. Phys.}\ }%
  \textbf{\bibinfo {volume} {64}},\ \bibinfo {pages} {39} (\bibinfo {year}
  {1996})%
  \bibAnnoteFile{NoStop}{ruby1996}%
\bibitem{mumford1960}%
  \BibitemOpen
  \bibfield{author}{%
  \bibinfo {author} {\bibfnamefont{W.}~\bibnamefont{Mumford}},\ }%
  \bibfield{journal}{%
  \bibinfo {journal} {Proceedings of the IRE}\ }%
  \textbf{\bibinfo {volume} {48}},\ \bibinfo {pages} {848} (\bibinfo {year}
  {1960})%
  \bibAnnoteFile{NoStop}{mumford1960}%
\bibitem{paul90}%
  \BibitemOpen
  \bibfield{author}{%
  \bibinfo {author} {\bibfnamefont{W.}~\bibnamefont{Paul}},\ }%
  \bibfield{journal}{%
  \bibinfo {journal} {Rev. of Mod. Phys.}\ }%
  \textbf{\bibinfo {volume} {62}},\ \bibinfo {pages} {531} (\bibinfo {year}
  {1990})%
  \bibAnnoteFile{NoStop}{paul90}%
\bibitem{raizen1992ionic}%
  \BibitemOpen
  \bibfield{author}{%
  \bibinfo {author} {\bibfnamefont{M.~G.}\ \bibnamefont{Raizen}}, \bibinfo
  {author} {\bibfnamefont{J.~M.}\ \bibnamefont{Gilligan}}, \bibinfo {author}
  {\bibfnamefont{J.~C.}\ \bibnamefont{Bergquist}}, \bibinfo {author}
  {\bibfnamefont{W.~M.}\ \bibnamefont{Itano}},\ and\ \bibinfo {author}
  {\bibfnamefont{D.~J.}\ \bibnamefont{Wineland}},\ }%
  \bibfield{journal}{%
  \bibinfo {journal} {Phys. Rev. A}\ }%
  \textbf{\bibinfo {volume} {45}},\ \bibinfo {pages} {6493} (\bibinfo {year}
  {1992})%
  \bibAnnoteFile{NoStop}{raizen1992ionic}%
\bibitem{drewsen1998large}%
  \BibitemOpen
  \bibfield{author}{%
  \bibinfo {author} {\bibfnamefont{M.}~\bibnamefont{Drewsen}}, \bibinfo
  {author} {\bibfnamefont{C.}~\bibnamefont{Brodersen}}, \bibinfo {author}
  {\bibfnamefont{L.}~\bibnamefont{Hornek{\ae}r}}, \bibinfo {author}
  {\bibfnamefont{J.~S.}\ \bibnamefont{Hangst}},\ and\ \bibinfo {author}
  {\bibfnamefont{J.~P.}\ \bibnamefont{Schifffer}},\ }%
  \bibfield{journal}{%
  \bibinfo {journal} {Phys. Rev. Lett.}\ }%
  \textbf{\bibinfo {volume} {81}},\ \bibinfo {pages} {2878} (\bibinfo {year}
  {1998})%
  \bibAnnoteFile{NoStop}{drewsen1998large}%
\bibitem{kielpinski2002architecture}%
  \BibitemOpen
  \bibfield{author}{%
  \bibinfo {author} {\bibfnamefont{D.}~\bibnamefont{Kielpinski}}, \bibinfo
  {author} {\bibfnamefont{C.}~\bibnamefont{Monroe}},\ and\ \bibinfo {author}
  {\bibfnamefont{D.~J.}\ \bibnamefont{Wineland}},\ }%
  \bibfield{journal}{%
  \bibinfo {journal} {Nature}\ }%
  \textbf{\bibinfo {volume} {417}},\ \bibinfo {pages} {709} (\bibinfo {year}
  {2002})%
  \bibAnnoteFile{NoStop}{kielpinski2002architecture}%
\bibitem{dougherty1996detection}%
  \BibitemOpen
  \bibfield{author}{%
  \bibinfo {author} {\bibfnamefont{W.~M.}\ \bibnamefont{Dougherty}}, \bibinfo
  {author} {\bibfnamefont{K.~J.}\ \bibnamefont{Bruland}}, \bibinfo {author}
  {\bibfnamefont{J.~L.}\ \bibnamefont{Garbini}},\ and\ \bibinfo {author}
  {\bibfnamefont{J.}~\bibnamefont{Sidles}},\ }%
  \bibfield{journal}{%
  \bibinfo {journal} {Meas. Sci. and Technol.}\ }%
  \textbf{\bibinfo {volume} {7}},\ \bibinfo {pages} {1733} (\bibinfo {year}
  {1996})%
  \bibAnnoteFile{NoStop}{dougherty1996detection}%
\bibitem{moreno2006parametric}%
  \BibitemOpen
  \bibfield{author}{%
  \bibinfo {author} {\bibfnamefont{M.}~\bibnamefont{Moreno-Moreno}}, \bibinfo
  {author} {\bibfnamefont{A.}~\bibnamefont{Raman}}, \bibinfo {author}
  {\bibfnamefont{J.}~\bibnamefont{Gomez-Herrero}},\ and\ \bibinfo {author}
  {\bibfnamefont{R.}~\bibnamefont{Reifenberger}},\ }%
  \bibfield{journal}{%
  \bibinfo {journal} {Appl. Phys. Lett.}\ }%
  \textbf{\bibinfo {volume} {88}},\ \bibinfo {pages} {193108} (\bibinfo {year}
  {2006})%
  \bibAnnoteFile{NoStop}{moreno2006parametric}%
\bibitem{requa06electromechanical}%
  \BibitemOpen
  \bibfield{author}{%
  \bibinfo {author} {\bibfnamefont{M.~V.}\ \bibnamefont{Requa}}\ and\ \bibinfo
  {author} {\bibfnamefont{K.~L.}\ \bibnamefont{Turner}},\ }%
  \bibfield{journal}{%
  \bibinfo {journal} {Appl. Phys. Lett.}\ }%
  \textbf{\bibinfo {volume} {88}},\ \bibinfo {pages} {263508} (\bibinfo {year}
  {2006})%
  \bibAnnoteFile{NoStop}{requa06electromechanical}%
\bibitem{turner98nature}%
  \BibitemOpen
  \bibfield{author}{%
  \bibinfo {author} {\bibfnamefont{K.~L.}\ \bibnamefont{Turner}}, \bibinfo
  {author} {\bibfnamefont{S.~A.}\ \bibnamefont{Miller}}, \bibinfo {author}
  {\bibfnamefont{P.~G.}\ \bibnamefont{Hartwell}}, \bibinfo {author}
  {\bibfnamefont{N.~C.}\ \bibnamefont{MacDonald}}, \bibinfo {author}
  {\bibfnamefont{S.~H.}\ \bibnamefont{Strogatz}},\ and\ \bibinfo {author}
  {\bibfnamefont{S.~G.}\ \bibnamefont{Adams}},\ }%
  \bibfield{journal}{%
  \bibinfo {journal} {Nature}\ }%
  \textbf{\bibinfo {volume} {396}},\ \bibinfo {pages} {149} (\bibinfo {year}
  {1998})%
  \bibAnnoteFile{NoStop}{turner98nature}%
\bibitem{eom2012nature}%
  \BibitemOpen
  \bibfield{author}{%
  \bibinfo {author} {\bibnamefont{{B.~H.~Eom, P.~K.~Day, H.~G.~LeDuc}}}\ and\
  \bibinfo {author} {\bibfnamefont{J.}~\bibnamefont{Zmuidzinas}},\ }%
  \bibfield{journal}{%
  \bibinfo {journal} {Nature Phys.}\ }%
  \textbf{\bibinfo {volume} {8}},\ \bibinfo {pages} {623–627} (\bibinfo
  {year} {2012})%
  \bibAnnoteFile{NoStop}{eom2012nature}%
\bibitem{szorkovszky2011}%
  \BibitemOpen
  \bibfield{author}{%
  \bibinfo {author} {\bibfnamefont{A.}~\bibnamefont{Szorkovszky}}, \bibinfo
  {author} {\bibfnamefont{A.~C.}\ \bibnamefont{Doherty}}, \bibinfo {author}
  {\bibfnamefont{G.~I.}\ \bibnamefont{Harris}},\ and\ \bibinfo {author}
  {\bibfnamefont{W.~P.}\ \bibnamefont{Bowen}},\ }%
  \bibfield{journal}{%
  \bibinfo {journal} {Physical Review Letters}\ }%
  \textbf{\bibinfo {volume} {107}},\ \bibinfo {pages} {213603} (\bibinfo {year}
  {2011})%
  \bibAnnoteFile{NoStop}{szorkovszky2011}%
\bibitem{binnig86}%
  \BibitemOpen
  \bibfield{author}{%
  \bibinfo {author} {\bibfnamefont{G.}~\bibnamefont{Binnig}}, \bibinfo {author}
  {\bibfnamefont{C.~F.}\ \bibnamefont{Quate}},\ and\ \bibinfo {author}
  {\bibfnamefont{C.}~\bibnamefont{Gerber}},\ }%
  \bibfield{journal}{%
  \bibinfo {journal} {Phys. Rev. Lett}\ }%
  \textbf{\bibinfo {volume} {56}},\ \bibinfo {pages} {930} (\bibinfo {year}
  {1986})%
  \bibAnnoteFile{NoStop}{binnig86}%
\bibitem{rugar91}%
  \BibitemOpen
  \bibfield{author}{%
  \bibinfo {author} {\bibfnamefont{D.}~\bibnamefont{Rugar}}\ and\ \bibinfo
  {author} {\bibfnamefont{P.}~\bibnamefont{Grutter}},\ }%
  \bibfield{journal}{%
  \bibinfo {journal} {Phys.~Rev.~Lett.}\ }%
  \textbf{\bibinfo {volume} {67}},\ \bibinfo {pages} {699} (\bibinfo {year}
  {1991})%
  \bibAnnoteFile{NoStop}{rugar91}%
\bibitem{difilippo92prl}%
  \BibitemOpen
  \bibfield{author}{%
  \bibinfo {author} {\bibfnamefont{F.}~\bibnamefont{DiFilippo}}, \bibinfo
  {author} {\bibfnamefont{V.}~\bibnamefont{Natarajan}}, \bibinfo {author}
  {\bibfnamefont{K.~R.}\ \bibnamefont{Boyce}},\ and\ \bibinfo {author}
  {\bibfnamefont{D.~E.}\ \bibnamefont{Pritchard}},\ }%
  \bibfield{journal}{%
  \bibinfo {journal} {Phys. Rev. Lett.}\ }%
  \textbf{\bibinfo {volume} {68}},\ \bibinfo {pages} {2859} (\bibinfo {year}
  {1992})%
  \bibAnnoteFile{NoStop}{difilippo92prl}%
\bibitem{natarajan95prl}%
  \BibitemOpen
  \bibfield{author}{%
  \bibinfo {author} {\bibfnamefont{V.}~\bibnamefont{Natarajan}}, \bibinfo
  {author} {\bibfnamefont{F.}~\bibnamefont{DiFilippo}},\ and\ \bibinfo {author}
  {\bibfnamefont{D.~E.}\ \bibnamefont{Pritchard}},\ }%
  \bibfield{journal}{%
  \bibinfo {journal} {Phys. Rev. Lett.}\ }%
  \textbf{\bibinfo {volume} {74}},\ \bibinfo {pages} {2855} (\bibinfo {year}
  {1995})%
  \bibAnnoteFile{NoStop}{natarajan95prl}%
\bibitem{almog2007noise}%
  \BibitemOpen
  \bibfield{author}{%
  \bibinfo {author} {\bibfnamefont{R.}~\bibnamefont{Almog}}, \bibinfo {author}
  {\bibfnamefont{S.}~\bibnamefont{Zaitsev}}, \bibinfo {author}
  {\bibfnamefont{O.}~\bibnamefont{Shtempluck}},\ and\ \bibinfo {author}
  {\bibfnamefont{E.}~\bibnamefont{Buks}},\ }%
  \bibfield{journal}{%
  \bibinfo {journal} {Phys.~Rev.~Lett.}\ }%
  \textbf{\bibinfo {volume} {98}},\ \bibinfo {pages} {078103} (\bibinfo {year}
  {2007})%
  \bibAnnoteFile{NoStop}{almog2007noise}%
\bibitem{batista2011cooling}%
  \BibitemOpen
  \bibfield{author}{%
  \bibinfo {author} {\bibfnamefont{A.~A.}\ \bibnamefont{Batista}},\ }%
  \bibfield{journal}{%
  \bibinfo {journal} {J. of Stat. Mech. (Theory and Experiment)}\ }%
  \textbf{\bibinfo {volume} {2011}},\ \bibinfo {pages} {P02007} (\bibinfo
  {year} {2011})%
  \bibAnnoteFile{NoStop}{batista2011cooling}%
\bibitem{verh96}%
  \BibitemOpen
  \bibfield{author}{%
  \bibinfo {author} {\bibfnamefont{F.}~\bibnamefont{Verhulst}},\ }%
  \emph{\bibinfo {title} {Nonlinear Differential Equations and Dynamical
  Systems}}\ (\bibinfo {publisher} {Springer-Verlag},\ \bibinfo {address} {New
  York},\ \bibinfo {year} {1996})%
  \bibAnnoteFile{NoStop}{verh96}%
\bibitem{Guck83}%
  \BibitemOpen
  \bibfield{author}{%
  \bibinfo {author} {\bibfnamefont{J.}~\bibnamefont{Guckenheimer}}\ and\
  \bibinfo {author} {\bibfnamefont{P.}~\bibnamefont{Holmes}},\ }%
  \emph{\bibinfo {title} {Nonlinear Oscillations, Dynamical Systems, and
  Bifurcations of Vector Fields}}\ (\bibinfo {publisher} {Springer-Verlag},\
  \bibinfo {address} {New York},\ \bibinfo {year} {1983})%
  \bibAnnoteFile{NoStop}{Guck83}%
\bibitem{hol81}%
  \BibitemOpen
  \bibfield{author}{%
  \bibinfo {author} {\bibfnamefont{C.}~\bibnamefont{Holmes}}\ and\ \bibinfo
  {author} {\bibfnamefont{P.}~\bibnamefont{Holmes}},\ }%
  \bibfield{journal}{%
  \bibinfo {journal} {J.~of Sound and Vib.}\ }%
  \textbf{\bibinfo {volume} {78}},\ \bibinfo {pages} {161} (\bibinfo {year}
  {1981})%
  \bibAnnoteFile{NoStop}{hol81}%
\bibitem{bat00}%
  \BibitemOpen
  \bibfield{author}{%
  \bibinfo {author} {\bibnamefont{{A.~A.~Batista, B.~Birnir,}}}\ and\ \bibinfo
  {author} {\bibnamefont{{M.~S.~Sherwin}}},\ }%
  \bibfield{journal}{%
  \bibinfo {journal} {Phys.~Rev.~B}\ }%
  \textbf{\bibinfo {volume} {61}},\ \bibinfo {pages} {15108} (\bibinfo {year}
  {2000})%
  \bibAnnoteFile{NoStop}{bat00}%
\bibitem{wies85}%
  \BibitemOpen
  \bibfield{author}{%
  \bibinfo {author} {\bibfnamefont{K.}~\bibnamefont{Wiesenfeld}}\ and\ \bibinfo
  {author} {\bibfnamefont{B.}~\bibnamefont{McNamara}},\ }%
  \bibfield{journal}{%
  \bibinfo {journal} {Phys.~Rev.~Lett.~}\ }%
  \textbf{\bibinfo {volume} {55}},\ \bibinfo {pages} {13} (\bibinfo {year}
  {1985})%
  \bibAnnoteFile{NoStop}{wies85}%
\bibitem{kubo66}%
  \BibitemOpen
  \bibfield{author}{%
  \bibinfo {author} {\bibfnamefont{R.}~\bibnamefont{Kubo}},\ }%
  \bibfield{journal}{%
  \bibinfo {journal} {Rep. Prog. Phys.}\ }%
  \textbf{\bibinfo {volume} {29}},\ \bibinfo {pages} {255} (\bibinfo {year}
  {1966})%
  \bibAnnoteFile{NoStop}{kubo66}%
\bibitem{batista2011signal}%
  \BibitemOpen
  \bibfield{author}{%
  \bibinfo {author} {\bibfnamefont{A.~A.}\ \bibnamefont{Batista}}\ and\
  \bibinfo {author} {\bibnamefont{{R.~S.~N. Moreira}}},\ }%
  \bibfield{journal}{%
  \bibinfo {journal} {Phys. Rev. E}\ }%
  \textbf{\bibinfo {volume} {84}},\ \bibinfo {pages} {061121} (\bibinfo {year}
  {2011})%
  \bibAnnoteFile{NoStop}{batista2011signal}%
\bibitem{ott1990controlling}%
  \BibitemOpen
  \bibfield{author}{%
  \bibinfo {author} {\bibfnamefont{E.}~\bibnamefont{Ott}}, \bibinfo {author}
  {\bibfnamefont{C.}~\bibnamefont{Grebogi}},\ and\ \bibinfo {author}
  {\bibfnamefont{J.}~\bibnamefont{Yorke}},\ }%
  \bibfield{journal}{%
  \bibinfo {journal} {Phys. Rev. Lett.}\ }%
  \textbf{\bibinfo {volume} {64}},\ \bibinfo {pages} {1196} (\bibinfo {year}
  {1990})%
  \bibAnnoteFile{NoStop}{ott1990controlling}%
\end{thebibliography}
%\bibliographystyle{apsrev4-1}   %>>>> makes bibtex use prbbib.bst

%Merlin.mbs v4.21 2009-07-09.
%
\newpage
\newpage

\end{document}